\documentclass[prd,twocolumn,groupedaddress,showpacs,nofootinbib]{revtex4}
\usepackage{graphicx}
\usepackage{dcolumn}
\usepackage{amssymb}
\usepackage{mathrsfs}
\usepackage{amsmath}
\usepackage{epsfig}
\usepackage[dvips]{color}
\usepackage{hhline}

\begin{document}

\title{Peccei-Quinn symmetry for Dirac seesaw and leptogenesis}

\author{Pei-Hong Gu}

\email{peihong.gu@sjtu.edu.cn}

\affiliation{Department of Physics and Astronomy, Shanghai Jiao Tong
University, 800 Dongchuan Road, Shanghai 200240, China}

\begin{abstract}

We extend the DFSZ invisible axion model to simultaneously explain small Dirac neutrino masses and cosmic matter-antimatter asymmetry. After the Peccei-Quinn and electroweak symmetry breaking, the effective Yukawa couplings of the Dirac neutrinos to the standard model Higgs scalar can be highly suppressed by the ratio of the vacuum expectation value of an iso-triplet Higgs scalar over the masses of some heavy gauge-singlet fermions, iso-doublet Higgs scalars or iso-triplet fermions. The iso-triplet fields can carry a zero or nonzero hypercharge. Through the decays of the heavy gauge-singlet fermions, iso-doublet scalars or iso-triplet fermions, we can obtain a lepton asymmetry in the left-handed leptons and an opposite lepton asymmetry in the right-handed neutrinos. Since the right-handed neutrinos do not participate in the sphaleron processes, the left-handed lepton asymmetry can be partially converted to a baryon asymmetry.

\end{abstract}

\pacs{98.80.Cq, 14.60.Pq, 14.80.Va}

\maketitle

\section{Introduction}

The phenomena of neutrino oscillations have been established by the atmospheric, solar, accelerator and reactor neutrino experiments \cite{olive2014}. This means three flavors of neutrinos should be massive and mixed. Since the neutrinos are massless in the $SU(3)_c^{}\times SU(2)_L^{}\times U(1)^{}_{Y}$ standard model (SM), we need new physics. Currently the most popular scheme for the neutrino mass generation is the so-called seesaw \cite{minkowski1977} mechanism which can highly suppress the neutrino masses by a small ratio of the electroweak scale over a newly high scale. Remarkably the neutrinos have a Majorana nature in the usual seesaw models \cite{minkowski1977,mw1980,flhj1989,ma1998,barr2003}. Such Majorana neutrino masses are induced by some lepton-number-violating interactions which can also generate a lepton asymmetry \cite{fy1986} and then give a baryon asymmetry in association with the sphaleron \cite{krs1985} processes. We hence can understand the cosmic matter-antimatter asymmetry which is the same as a baryon asymmetry. This baryogensis scenario in the lepton-number-violating seesaw context is the well known leptogenesis \cite{fy1986} mechanism and has been widely studied
\cite{lpy1986,fps1995,ms1998,bcst1999,hambye2001,di2002,gnrrs2003,hs2004,bbp2005,ma2006,dnn2008,dhh2014,ksy2015,fmmn2015}.

However, one should keep in mind that the theoretical assumption of the lepton number violation and then the Majorana neutrinos have not been confirmed by any experiments. So it is worth studying the Dirac neutrinos \cite{rw1983,dlrw1999,mp2002,gh2006,tt2006,dsz2007,gu2012}. In particular, we can construct some lepton-number-conserving Dirac seesaw models \cite{rw1983,mp2002,gh2006,gu2012} to generate the small Dirac neutrino masses. The key of the Dirac seesaw models is that the effective Yukawa couplings of the right-handed neutrinos to the SM leptons and Higgs scalar can be suppressed by a ratio of ceratin symmetry breaking scale over some heavy field masses. Through the out-of-equilibrium and CP-violating decays of these heavy fields, we can obtain a lepton asymmetry in the SM left-handed leptons and an opposite lepton asymmetry in the right-handed neutrinos although the lepton number is totally zero \cite{dlrw1999,mp2002,gh2006,tt2006}. The right-handed neutrinos then will go into equilibrium with the left-handed neutrinos at a very low temperature where the sphalerons have already stopped working. Therefore, the sphalerons can partially convert the induced lepton asymmetry in the SM leptons to a baryon asymmetry. This type of leptogenesis is named as the neutrinogenesis \cite{dlrw1999} mechanism.

The SM encounters other challenges besides the small neutrino masses and the cosmic baryon asymmetry. In order to solve those problems, people have also extended the SM in other ways except for the seesaw scenario. For example, the invisible axion models \cite{kim1979,dfs1981} based on the Peccei-Quinn (PQ) symmetry \cite{pq1977,weinberg1978,wilczek1978} have been studied widely by theorists and experimentalists since they can solve the strong CP problem. Due to the unobserved axion, the PQ symmetry breaking scale now has a low limit far above the electroweak scale \cite{olive2014}. Furthermore, for a proper choice of the breaking scale of the PQ symmetry and the initial value of the strong CP phase, the invisible axion can account for the dark matter relic density in the universe \cite{olive2014}. In some interesting models for the neutrino mass generation, the PQ symmetry also plays an essential role \cite{shin1987}.

We would like to point out the usual Dirac seesaw models contain an arbitrary breaking scale of the additional discrete, global or gauge symmetry. To fix or constrain this symmetry breaking scale, we can connect it to other new physics. For example, in a class of mirror models \cite{gu2012}, the additional symmetry is a mirror electroweak symmetry so that it can be fixed by the dark matter mass.

In this paper we shall make use of the PQ symmetry to forbid the Yukawa couplings of the right-handed neutrinos to the SM leptons and Higgs scalar. Specifically we shall extend the DFSZ \cite{dfs1981} invisible axion model by three gauge-singlet right-handed neutrinos, an iso-triplet Higgs scalar with or without hypercharge, as well as some heavy gauge-singlet fermions, iso-doublet Higgs scalars or iso-triplet fermions. After the PQ and electroweak symmetry breaking, the iso-triplet Higgs scalar can acquire an induced vacuum expectation value (VEV) constrained by the $\rho$ parameter. This VEV can help us to naturally suppress the Dirac neutrino masses by its ratio over the masses of the heavy gauge-singlet fermions, iso-doublet Higgs scalars or iso-triplet fermions. Meanwhile, the decays of the heavy gauge-singlet fermions, iso-doublet Higgs scalars or iso-triplet fermions can realize a neutrinogenesis to explain the cosmic matter-antimatter asymmetry.

\section{The DFSZ model}

Before introducing our models, we briefly review the DFSZ invisible axion model which contains three generations of fermions,
\begin{eqnarray}
\!\!\!\!\!\!&&q_L^{}(\!\!\begin{array}{c}3,2,+\frac{1}{6})(0\end{array}\!\!)=\left[\begin{array}{c}u_L^{}\\
[2mm]
d_L^{}\end{array}\right]\!,
~~l_L^{}(\!\!\begin{array}{c}1,2,-\frac{1}{2})(0\end{array}\!\!)=\left[\begin{array}{c}\nu_L^{}\\
[2mm]
e_L^{}\end{array}\right]\!,\nonumber\\
[2mm]
\!\!\!\!\!\!&&u_R^{}(\!\!\begin{array}{c}1,1,+\frac{2}{3})(+1\end{array}\!\!),~~ d_R^{}(\!\!\begin{array}{c}1,1,-\frac{1}{3})(+1\end{array}\!\!),
~~ e_R^{}(\!\!\begin{array}{c}1,1,-1)(+1\end{array}\!\!),\nonumber\\
\!\!\!\!\!\!&&
\end{eqnarray}
as well as three Higgs scalars,
\begin{eqnarray}
\!\!\!\!\!\!&&\phi_1^{}(\!\!\begin{array}{c}1,2,-\frac{1}{2})(-1\end{array}\!\!)=\left[\begin{array}{c}\phi_1^{0}\\
[2mm]
\phi_1^{-}\end{array}\right]\!,~ \phi_2^{}(\!\!\begin{array}{c}1,2,-\frac{1}{2})(+1\end{array}\!\!)=\left[\begin{array}{c}\phi_2^{0}\\
[2mm]
\phi_2^{-}\end{array}\right]\!,\nonumber\\
\!\!\!\!\!\!&&\chi(\!\!\begin{array}{c}1,1,0)(+1\end{array}\!\!).
\end{eqnarray}
Here and thereafter the first brackets following the fields describe the transformations under the $SU(3)_c^{}\times SU(2)_L^{}\times U(1)^{}_{Y}$ gauge groups while the second ones denote the charges under a $U(1)^{}_{\textrm{PQ}}$ global symmetry.

We write down the kinetic terms of the above fermions and scalars,
\begin{eqnarray}
\label{kinetic}
\mathcal{L}_K^{}&=&i\bar{q}_L^{}\gamma^\mu_{}D_\mu^{}q_L^{}+i\bar{u}_R^{}\gamma^\mu_{}D_\mu^{}u_R^{}
+i\bar{d}_R^{}\gamma^\mu_{}D_\mu^{}d_R^{}\nonumber\\
&&+i\bar{l}_L^{}\gamma^\mu_{}D_\mu^{}l_L^{}+i\bar{e}_R^{}\gamma^\mu_{}D_\mu^{}e_R^{}+(D_\mu^{}\phi_1^{})^\dagger_{}D^\mu_{}\phi_1^{}\nonumber\\
&&
+(D_\mu^{}\phi_2^{})^\dagger_{}D^\mu_{}\phi_2^{}+(\partial_\mu^{}\chi)^\dagger_{}\partial^\mu_{}\chi\,,
\end{eqnarray}
with the covariant derivatives,
\begin{eqnarray}
D_\mu^{}q_L^{}&=&\left(\partial_\mu^{}-ig_3^{}\frac{\lambda_a^{}}{2}G^a_{\mu}-ig\frac{\tau_a^{}}{2}W^a_\mu-i\frac{1}{6}g'B^{}_\mu\right)q_L^{}\,,\nonumber\\
D_\mu^{}u_R^{}&=&\left(\partial_\mu^{}-ig_3^{}\frac{\lambda_a^{}}{2}G^a_{\mu}-i\frac{2}{3}g'B^{}_\mu\right)u_R^{}\,,\nonumber\\
D_\mu^{}d_R^{}&=&\left(\partial_\mu^{}-ig_3^{}\frac{\lambda_a^{}}{2}G^a_{\mu}+i\frac{1}{3}g'B^{}_\mu\right)d_R^{}\,,\nonumber\\
D_\mu^{}l_L^{}&=&\left(\partial_\mu^{}-ig\frac{\tau_a^{}}{2}W^a_\mu+i\frac{1}{2}g'B^{}_\mu\right)l_L^{}\,,\nonumber\\
D_\mu^{}e_R^{}&=&\left(\partial_\mu^{}+ig'B^{}_\mu\right)e_R^{}\,,\nonumber\\
D_\mu^{}\phi_1^{}&=&\left(\partial_\mu^{}-ig\frac{\tau_a^{}}{2}W^a_\mu+i\frac{1}{2}g'B^{}_\mu\right)\phi_1^{}\,,\nonumber\\
D_\mu^{}\phi_2^{}&=&\left(\partial_\mu^{}-ig\frac{\tau_a^{}}{2}W^a_\mu+i\frac{1}{2}g'B^{}_\mu\right)\phi_2^{}\,.
\end{eqnarray}
Here $g_3^{}$, $g$ and $g'$ are the $SU(3)_c^{}$, $SU(2)_L^{}$ and $U(1)_Y^{}$ gauge couplings, $G^a_{\mu}(a=1,2,...,8)$, $W^a_{\mu}(a=1,2,3)$ and $B_\mu^{}$ are the corresponding gauge fields, while $\lambda_a^{}(a=1,2,...,8)$ and $\tau_a^{}(a=1,2,3)$ are the Gell-Mann and Pauli matrices. Under the $SU(3)_c^{}\times SU(2)_L^{}\times U(1)^{}_{Y}$ and $U(1)^{}_{\textrm{PQ}}$ symmetries, we can give the Yukawa interactions,
\begin{eqnarray}
\label{yukawa}
\mathcal{L}_Y^{}=-y_u^{}\bar{q}_L^{}\phi_1^{}u_R^{}-y_d^{}\bar{q}_L^{}\tilde{\phi}_2^{}d_R^{}
-y_e^{}\bar{l}_L^{}\tilde{\phi}_2^{}e_R^{}+\textrm{H.c.}\,,
\end{eqnarray}
and the scalar potential,
\begin{eqnarray}
\label{potential1}
V(\chi,\phi_1^{},\phi_2^{})&=&\mu_1^2\phi_1^\dagger\phi_1^{}
+\mu_2^2\phi_2^\dagger\phi_2^{}+\mu_3^2\chi^\dagger_{}\chi+\lambda_1^{}(\phi_1^\dagger\phi_1^{})^2_{}\nonumber\\
&&
+\lambda_2^{}(\phi_2^\dagger\phi_2^{})^2_{}
+\lambda_3^{}(\chi^\dagger_{}\chi)^2_{}+\lambda_{4}^{}\phi_1^\dagger\phi_1^{}\phi_2^\dagger\phi_2^{}
\nonumber\\
&&
+\lambda_{5}^{}\phi_1^\dagger\phi_2^{}\phi_2^\dagger\phi_1^{}
+\lambda_{6}^{}\phi_1^\dagger\phi_1^{}\chi_{}^\dagger\chi\nonumber\\
&&+\lambda_{7}^{}\phi_2^\dagger\phi_2^{}\chi_{}^\dagger\chi
+\lambda_8^{}(\chi^2_{}\phi_2^\dagger\phi_1^{}+\textrm{H.c.})\,.
\end{eqnarray}

After the gauge-singlet scalar $\chi$ develops a VEV,
\begin{eqnarray}
\langle\chi\rangle=\frac{1}{\sqrt{2}}f_{\textrm{PQ}}^{}\,,
\end{eqnarray}
to spontaneously break the $U(1)^{}_{\textrm{PQ}}$ global symmetry, it can be rewritten by
\begin{eqnarray}
\chi=\frac{1}{\sqrt{2}}(f_{\textrm{PQ}}^{}+h_{\textrm{PQ}}^{})\exp\left(i\frac{a}{f_{\textrm{PQ}}^{}}\right)\,,
\end{eqnarray}
where $h_{\textrm{PQ}}^{}$ is a massive Higgs boson while $a$ is a Nambu-Goldstone
boson. By making the following phase rotation,
\begin{eqnarray}
&&\phi_{1}^{}\exp\left(i\frac{a}{f_{\textrm{PQ}}^{}}\right)\rightarrow \phi_{1}^{}\,,~~\phi_{2}^{}\exp\left(-i\frac{a}{f_{\textrm{PQ}}^{}}\right)\rightarrow \phi_{2}^{}\,,\nonumber\\
&&u_R^{}\exp\left(-i\frac{a}{f_{\textrm{PQ}}^{}}\right)\rightarrow u_R^{}\,,~~
d_R^{}\exp\left(-i\frac{a}{f_{\textrm{PQ}}^{}}\right)\rightarrow d_R^{}\,,\nonumber\\
&&e_R^{}\exp\left(-i\frac{a}{f_{\textrm{PQ}}^{}}\right)\rightarrow e_R^{}\,,
\end{eqnarray}
the kinetic terms (\ref{kinetic}) can give us the axial couplings of the Nambu-Goldstone
boson $a$ to the SM fermions $u$, $d$ and $e$,
\begin{eqnarray}
\mathcal{L}&\supset&-\frac{\partial_\mu^{}a}{f_{\textrm{PQ}}^{}}(\bar{u}_R^{}\gamma^\mu_{}u_R^{}+\bar{d}_R^{}\gamma^\mu_{}d_R^{}
+\bar{e}_R^{}\gamma^\mu_{}e_R^{})\nonumber\\
&\supset&-\frac{\partial_\mu^{}a}{2f_{\textrm{PQ}}^{}}(\bar{u}\gamma^\mu_{}\gamma_5^{}u+\bar{d}\gamma^\mu_{}\gamma_5^{}d+\bar{e}\gamma^\mu_{}\gamma_5^{}e)\,.
\end{eqnarray}
The non-perturbative QCD Lagrangian then should be
\begin{eqnarray}
\mathcal{L}_{\textrm{QCD}}^{}\supset-\bar{\theta}\frac{g^2_3}{32\pi^2_{}}G\tilde{G}~~\textrm{with}~~
\bar{\theta}=\theta+\frac{a}{2f_{\textrm{PQ}}^{}}\,,
\end{eqnarray}
where $\theta$ is a constant from the quark mass matrices and the QCD $\Theta$-vacuum. Clearly, the physical strong CP phase $\bar{\theta}$ now can naturally roll into a tiny value to solve the strong CP problem since it now has become a dynamical field. Therefore, the global symmetry $U(1)^{}_{\textrm{PQ}}$ is the PQ symmetry while the Nambu-Goldstone
boson $a$ is the axion. The PQ symmetry should
be broken at a high scale $f_{\textrm{PQ}}^{}\gtrsim 10^{10}_{}\,\textrm{GeV}$ to fulfill the
experimental constraints \cite{olive2014}. From the color anomaly the axion can pick up a tiny mass. For an appropriate PQ symmetry breaking scale $f_{\textrm{PQ}}^{} \lesssim 10^{12}_{}\,\textrm{GeV}$, the axion can serve as a cold dark matter particle if the strong CP phase $\bar{\Theta}$ has an initial value of the order of $\mathcal{O}(1)$ \cite{olive2014}.

The $[SU(2)_L^{}]$-doublet Higgs scalars $\phi_{1,2}^{}$ are responsible for the spontaneous electroweak symmetry breaking. Their VEVs should be
\begin{eqnarray}
&&\langle\phi_1^{}\rangle=\left[\begin{array}{c}\langle\phi_1^{0}\rangle\\
[2mm]
0\end{array}\right]\!,~ \langle\phi_2^{}\rangle=\left[\begin{array}{c}\langle\phi_2^{0}\rangle\\
[2mm]
0\end{array}\right]~~\textrm{with}\nonumber\\
&&\tan\beta=\frac{\langle\phi_1^{0}\rangle}{\langle\phi_2^{0}\rangle}\,.
\end{eqnarray}
We can conveniently define
\begin{eqnarray}
\label{phi}
\phi=\frac{\langle\phi_{1}^{0}\rangle\phi_{1}^{}+\langle\phi_{2}^{0}\rangle\phi_{2}^{}}{\sqrt{\langle\phi_{1}^{0}\rangle^2_{}+\langle\phi_{2}^{0}\rangle^2_{}}}\,,~~
\phi'=\frac{\langle\phi_{2}^{0}\rangle\phi_{1}^{}-\langle\phi_{1}^{0}\rangle\phi_{2}^{}}{\sqrt{\langle\phi_{1}^{0}\rangle^2_{}+\langle\phi_{2}^{0}\rangle^2_{}}}\,,
\end{eqnarray}
and then obtain
\begin{eqnarray}
\langle\phi\rangle&=&\left[\begin{array}{c}\langle\phi_{}^{0}\rangle\\
[2mm]
0\end{array}\right]~~\textrm{with}~~\langle\phi^0_{}\rangle=\sqrt{\langle\phi_{1}^{0}\rangle^2_{}+\langle\phi_{2}^{0}\rangle^2_{}}\,,\nonumber\\
\langle\phi'\rangle&=&0\,.
\end{eqnarray}
This means the newly defined $\phi$ will drive the electroweak symmetry breaking. It is easy to see the perturbation requirement in the Yukawa interactions can constrain the rotation angle $\beta$ by
\begin{eqnarray}
\frac{1}{\sqrt{\frac{4\pi \langle\phi^0_{}\rangle^2_{}}{m_t^2}-1}}<\tan\beta < \sqrt{\frac{4\pi \langle\phi^0_{}\rangle^2_{}}{m_b^2}-1}\,.
\end{eqnarray}
By inputting \cite{olive2014}
\begin{eqnarray}
m_t^{}=173\,\textrm{GeV}\,,~~m_b^{}=4.18\,\textrm{GeV}\,,~~\langle\phi^0_{}\rangle= 174\,\textrm{GeV}\,,~~
\end{eqnarray}
we can read
\begin{eqnarray}
0.3\lesssim\tan\beta\lesssim147\,.
\end{eqnarray}

\section{Higgs triplets and right-handed neutrinos}

We now introduce the Higgs triplets with or without hypercharge,
\begin{subequations}
\begin{eqnarray}
\Sigma(\!\!\begin{array}{c}1,3,0)(+2\end{array}\!\!)
&=&\left[\begin{array}{cc}\frac{1}{\sqrt{2}}\sigma_{}^{0}&\sigma_{2}^{+}\\
[2mm]
\sigma_1^{-}&-\frac{1}{\sqrt{2}}\sigma_{}^{0}\end{array}\right]\!;\\
[2mm]\Delta(\!\!\begin{array}{c}1,3,+1)(+2\end{array}\!\!)&=&
\left[\begin{array}{cc}\frac{1}{\sqrt{2}}\delta_{}^{+}&\delta_{}^{++}\\
[2mm]
\delta_{}^{0}&-\frac{1}{\sqrt{2}}\delta_{}^{+}\end{array}\right]\!,
\end{eqnarray}
\end{subequations}
which have the kinetic terms as below,
\begin{subequations}
\begin{eqnarray}
\mathcal{L}_K^{}&\supset&\textrm{Tr}[(D_\mu^{}\Sigma)^\dagger_{}D^\mu_{}\Sigma]~~\textrm{with}\nonumber\\
&&D_\mu^{}\Sigma=\partial_\mu^{}\Sigma -ig\left[\frac{\tau_a^{}}{2}W^a_\mu,\Sigma\right]\,;\\
[2mm]
\mathcal{L}_K^{}&\supset&\textrm{Tr}[(D_\mu^{}\Delta)^\dagger_{}D^\mu_{}\Delta]~~\textrm{with}\nonumber\\
&&D_\mu^{}\Delta=\partial_\mu^{}\Delta -ig\left[\frac{\tau_a^{}}{2}W^a_\mu,\Delta\right]-ig'B_\mu^{}\Delta\,.~~~~
\end{eqnarray}
\end{subequations}
The supplement of the potential (\ref{potential1}) should be
\begin{subequations}
\begin{eqnarray}
V(\Sigma)&=&(\mu_\Sigma^2 +\zeta_1^{}\phi_1^{\dagger}\phi_1^{}
+\zeta_2^{}\phi_2^{\dagger}\phi_2^{}+\zeta_3^{}\chi_{}^{\dagger}\chi)\textrm{Tr}(\Sigma^\dagger_{}\Sigma)\nonumber\\
&&+
\zeta_4^{}[\textrm{Tr}(\Sigma^\dagger_{}\Sigma)]^2_{}
+\zeta_5^{}\textrm{Tr}(\Sigma^\dagger_{}\Sigma^\dagger_{})\textrm{Tr}(\Sigma\Sigma))\nonumber\\
&&+
\zeta_6^{}\textrm{Tr}[(\Sigma^\dagger_{}\Sigma)^2_{}]
+\zeta_7^{}\textrm{Tr}(\Sigma^\dagger_{}\Sigma^\dagger_{}\Sigma\Sigma)\nonumber\\
&&
+\omega_{\Sigma}^{}(\phi_1^T i\tau_2^{}\Sigma \tilde{\phi}_2^{}+\textrm{H.c.})\,;\\
[2mm]
V(\Delta)&=&(\mu_\Delta^2 +\zeta_1^{}\phi_1^{\dagger}\phi_1^{}+\zeta_2^{}\phi_2^{\dagger}\phi_2^{}
+\zeta_3^{}\chi_{}^{\dagger}\chi)\textrm{Tr}(\Delta^\dagger_{}\Delta)\nonumber\\
&&+
\zeta_4^{}[\textrm{Tr}(\Delta^\dagger_{}\Delta)]^2_{}
+\zeta_5^{}\textrm{Tr}(\Delta^\dagger_{}\Delta^\dagger_{})\textrm{Tr}(\Delta\Delta)\nonumber\\
&&+
\zeta_6^{}\textrm{Tr}[(\Delta^\dagger_{}\Delta)^2_{}]
+\zeta_7^{}\textrm{Tr}(\Delta^\dagger_{}\Delta^\dagger_{}\Delta\Delta)\nonumber\\
&&
+\omega_{\Delta}^{}(\phi_1^T i\tau_2^{}\Delta \phi_1^{}+\textrm{H.c.})\,.
\end{eqnarray}
\end{subequations}

After the Higgs doublets $\phi_{1,2}^{}$ develop their VEVs for the electroweak symmetry breaking, the Higgs triplets can acquire the induced VEVs,
\begin{subequations}
\begin{eqnarray}
\langle\Sigma\rangle&=&\left[\begin{array}{cc}\frac{\langle\sigma_{}^{0}\rangle}{\sqrt{2}}&0\\
[2mm]
0&-\frac{\langle\sigma_{}^{0}\rangle}{\sqrt{2}}\end{array}\right]~~\textrm{with}\nonumber\\
[2mm]
&&\langle\sigma_{}^{0}\rangle\simeq -\frac{\omega_\Sigma\langle\phi_1^{0}\rangle\langle\phi_2^{0}\rangle}{\sqrt{2}M_{\Sigma}^2}
=-\frac{\omega_\Sigma\langle\phi\rangle^2_{}\sin2\beta}{2\sqrt{2}M_{\Sigma}^2}\,;\nonumber\\
&&\\
\langle\Delta\rangle&=&\left[\begin{array}{cc}0&0\\
[2mm]
\langle\delta_{}^{0}\rangle&0\end{array}\right]~~\textrm{with}\nonumber\\
[2mm]
&&\langle\delta_{}^{0}\rangle\simeq -\frac{\omega_\Delta\langle\phi_1^{0}\rangle^2_{}}{M_{\Delta}^2}=-\frac{\omega_\Delta\langle\phi\rangle^2_{}\sin^2_{}\beta}{M_{\Delta}^2}\,,
\end{eqnarray}
\end{subequations}
where the Higgs triplet masses $M_{\Sigma,\Delta}^2$ have been given by
\begin{subequations}
\begin{eqnarray}
M_\Sigma^2&=&\mu_\Sigma^2 +\zeta_1^{}\langle\phi_1^{0}\rangle^2_{}
+\zeta_2^{}\langle\phi_2^{0}\rangle^2_{}+\zeta_3^{}\langle\chi\rangle^2_{}\,;\\
[2mm]
M_\Delta^2&=&\mu_\Delta^2 +\zeta_1^{}\langle\phi_1^{0}\rangle^2_{}
+\zeta_2^{}\langle\phi_2^{0}\rangle^2_{}+\zeta_3^{}\langle\chi\rangle^2_{}\,.
\end{eqnarray}
\end{subequations}
It is well known the VEV of a Higgs triplet will affect the $\rho$ parameter \cite{olive2014},
\begin{eqnarray}
\label{limit}
\rho=\frac{M_W^2}{M_Z^2\cos^2_{}\theta_W^{}}=1.00040\pm 0.00024\,.
\end{eqnarray}
In the presence of two Higgs doublets $\phi_{1,2}^{}$ and a Higgs triplet $\Sigma$ or $\Delta$, we can express the $\rho$ parameter by
\begin{subequations}
\begin{eqnarray}
\rho&=&\frac{\langle\phi^0_{}\rangle^2_{}+4\langle\sigma^0_{}\rangle^2_{}}
{\langle\phi^0_{}\rangle^2_{}}\,;\\
[2mm]
\rho&=&
\frac{\langle\phi^0_{}\rangle^2_{}+2\langle\delta^0_{}\rangle^2_{}}
{\langle\phi^0_{}\rangle^2_{}+4\langle\delta^0_{}\rangle^2_{}}\,.
\end{eqnarray}
\end{subequations}
By inserting
\begin{subequations}
\begin{eqnarray}
\sqrt{\langle\phi^0_{}\rangle^2_{}+4\langle\delta^0_{}\rangle^2_{}}&=&174\,\textrm{GeV}\,,~~0.99968\leq \rho\leq 1.00112\,;\nonumber\\
&&\\
\sqrt{\langle\phi^0_{}\rangle^2_{}+2\langle\delta^0_{}\rangle^2_{}}&=&174\,\textrm{GeV}\,,~~0.99968\leq \rho\leq 1.00112\,,\nonumber\\
&&
\end{eqnarray}
\end{subequations}
we can derive the upper bounds on the VEVs of the Higgs triplets,
\begin{subequations}
\begin{eqnarray}
\langle\sigma^0_{}\rangle&\leq& 2.9\,\textrm{GeV}\,;\\
[2mm]
\langle\delta^0_{}\rangle&\leq& 2.2\,\textrm{GeV}\,.
\end{eqnarray}
\end{subequations}

Our models also contain three right-handed neutrinos, which are the $SU(3)_c^{}\times SU(2)_L^{}\times U(1)^{}_{Y}$ singlets but carry a $U(1)^{}_{\textrm{PQ}}$ charge as below,
\begin{eqnarray}
\nu_R^{}(\!\!\begin{array}{c}1,1,0)(+4\end{array}\!\!).
\end{eqnarray}
Therefore, the right-handed neutrinos are forbidden to have the following gauge-invariant Yukawa couplings and Majorana masses, i.e.
\begin{eqnarray}
\mathcal{L}&\supset\!\!\!\!\!/&-y_{\phi_1}^{}\bar{l}_L^{} \phi_1^{}\nu_R^{}-y_{\phi_2}^{}\bar{l}_L^{} \phi_2^{}\nu_R^{}-\frac{1}{2}m_{\nu_R}^{}\bar{\nu}_R^{}\nu_R^c+\textrm{H.c.}\,,\nonumber\\
&&
\end{eqnarray}
except for their kinetic terms,
\begin{eqnarray}
\mathcal{L}\supset i \bar{\nu}_R^{}\gamma^\mu_{}\partial_\mu^{}\nu_R^{}\,.
\end{eqnarray}
Meanwhile, the gauge-invariant Yukawa couplings of the Higgs triplet with hypercharge to the lepton doublets are also absent from the Lagrangian due to the PQ symmetry, i.e.
\begin{eqnarray}
\mathcal{L}\supset\!\!\!\!\!/~-\frac{1}{2}f_\Delta^{}\bar{l}_L^c i\tau_2^{}\Delta l_L^{}+\textrm{H.c.}\,.
\end{eqnarray}
In consequence, the neutrinos should keep massless in the present context.

\section{Dirac seesaw models}

In this section we will draw the outline of our models with the heavy fermion singlets, the heavy Higgs doublets or the heavy fermion triplets. The generation of the neutrino masses and the baryon asymmetry will be discussed in the later sections. According to the usual type-I, II and III seesaw models for the Majorana neutrinos, we would like to name our models with the heavy fermion singlets, the heavy Higgs doublets and the heavy fermion triplets as the type-I, type-II and type-III Dirac seesaw, respectively.

\subsection{Type-I Dirac seesaw}

The type-I Dirac seesaw contains the gauge-singlet fermions and scalar as follows,
\begin{eqnarray}
N_R^{}(\!\!\begin{array}{c}1,1,0)(+1\end{array}\!\!),~~N'^{}_R(\!\!\begin{array}{c}1,1,0)(0\end{array}\!\!),~~\omega(\!\!\begin{array}{c}1,1,0)(-4\end{array}\!\!).
\end{eqnarray}
The allowed kinetic, Yukawa and scalar interactions are
\begin{subequations}
\label{lag1}
\begin{eqnarray}
\mathcal{L}&\supset& i\bar{N}_R^{}\gamma^\mu_{}D_\mu^{}N_R^{}+i\bar{N}'^{}_R\gamma^\mu_{}D_\mu^{}N'^{}_R+(\partial_\mu^{}\omega)^\dagger_{}\partial^\mu_{}\omega
\nonumber\\
&&-y_{N}^{}\bar{l}_L^{}\phi_1^{}N^{}_R-y_{N'}^{}\omega\bar{N'}^c_R \nu_R^{}
-f_{N}^{}\chi\bar{N}_R^{}N'^{c}_R \nonumber\\
&&-\kappa_{\omega\Sigma}^{}\omega\phi_2^T i\tau_2^{} \Sigma \tilde{\phi}_1^{}+\textrm{H.c.}-[\mu_\omega^2+\alpha_1^{}\phi_1^\dagger\phi_1^{}+\alpha_2^{}\phi_2^\dagger\phi_2^{}\nonumber\\
&&+\alpha_3^{}\chi^\dagger_{}\chi
+\alpha_4^{}\textrm{Tr}(\Sigma^\dagger_{}\Sigma)]\omega^\dagger_{}\omega-\alpha_5^{}(\omega^\dagger_{}\omega)^2_{}\,;\\
[2mm]
\mathcal{L}&\supset& i\bar{N}_R^{}\gamma^\mu_{}D_\mu^{}N_R^{}+i\bar{N}'^{}_R\gamma^\mu_{}D_\mu^{}N'^{}_R+(\partial_\mu^{}\omega)^\dagger_{}\partial^\mu_{}\omega
\nonumber\\
&&-y_{N}^{}\bar{l}_L^{}\phi_1^{}N^{}_R-y_{N'}^{}\omega\bar{N'}^c_R \nu_R^{}
-f_{N}^{}\chi\bar{N}_R^{}N'^{c}_R \nonumber\\
&&-\kappa_{\omega\Delta}^{}\omega\phi_2^T i\tau_2^{} \Delta \phi_2^{}+\textrm{H.c.}-[\mu_\omega^2+\alpha_1^{}\phi_1^\dagger\phi_1^{}+\alpha_2^{}\phi_2^\dagger\phi_2^{}\nonumber\\
&&+\alpha_3^{}\chi^\dagger_{}\chi
+\alpha_4^{}\textrm{Tr}(\Delta^\dagger_{}\Delta)]\omega^\dagger_{}\omega-\alpha_5^{}(\omega^\dagger_{}\omega)^2_{}\,.
\end{eqnarray}
\end{subequations}
Here we have prevented the fermion singlets $N'^{}_R$ from the gauge-invariant Majorana masses by imposing a conserved global symmetry of lepton number, under which the singlet fermions $N_{R}^{}$ and $N'^{c}_R$, the right-handed neutrinos $\nu_R^{}$ and the SM leptons $l_L^{}$ and $e_R^{}$ all carry a lepton number of one unit.

After the PQ symmetry breaking, we can obtain a mass term between the fermion singlets $N_{R}^{}$ and $N'^{}_R$, i.e.
\begin{eqnarray}
\mathcal{L}\supset-M_{N}^{}\bar{N}_R^{}N'^c_R +\textrm{H.c.}
~~\textrm{with}~~M_N^{}=f_N^{}\langle\chi\rangle\,.
\end{eqnarray}
Without loss of generality, it is convenient to choose a basis where the masses of the fermion singlets are real and diagonal, i.e.
\begin{eqnarray}
M_N^{}&=&\textrm{diag}\{M_{N_1}^{},M_{N_2}^{},...\}\,,
\end{eqnarray}
and then define the following vector-like fermions,
\begin{eqnarray}
N^{}_a=N'^{c}_{Ra}+N^{}_{Ra}\,.
\end{eqnarray}
As for the scalar singlet $\omega$, its mass is dominated by
\begin{eqnarray}
M_\omega^2\!&\simeq& \!\mu_\omega^2+\alpha_1^{}\langle\phi_1^{}\rangle^2_{}+\alpha_2^{}\langle\phi_2^{}\rangle^2_{}+\alpha_3^{}\langle\chi\rangle^2_{} \gg \langle\Sigma\rangle^2_{}, \langle\Delta\rangle^2_{}\,.\nonumber\\
&&
\end{eqnarray}

\subsection{Type-II Dirac seesaw}

The heavy Higgs doublets for the type-II Dirac seesaw are denoted by
\begin{eqnarray}
\eta(\!\!\begin{array}{c}1,2,-\frac{1}{2})(-4\end{array}\!\!)=\left[\begin{array}{c}\eta_{}^{0}\\
[2mm]
\eta_{}^{-}\end{array}\right]\!,
\end{eqnarray}
which have the kinetic, Yukawa and scalar interactions,
\begin{subequations}
\label{lag2}
\begin{eqnarray}
\mathcal{L}&\supset& (D_\mu^{}\eta)^\dagger_{}D^\mu_{}\eta-y_\eta^{}\bar{l}_L^{}\eta\nu_R^{}-\kappa_{\eta\Sigma}^{}\chi
\tilde{\phi}_{1}^{T} i\tau_2^{}\Sigma \eta+\textrm{H.c.}\nonumber\\
&&-[\mu_\eta^2+\beta_1^{}\phi_1^\dagger\phi_1^{}+\beta_2^{}\phi_2^\dagger\phi_2^{}+\beta_3^{}\chi^\dagger_{}\chi
\nonumber\\
&&+\beta_4^{}\textrm{Tr}(\Sigma^\dagger_{}\Sigma)]\eta^\dagger_{}\eta-\beta_5^{}(\eta^\dagger_{}\eta)^2_{}\,;\\
[2mm]
\mathcal{L}&\supset& (D_\mu^{}\eta)^\dagger_{}D^\mu_{}\eta-y_\eta^{}\bar{l}_L^{}\eta\nu_R^{}-\kappa_{\eta\Delta}^{}\chi\phi^T_{2} i\tau_2^{}\Delta \eta+\textrm{H.c.}\nonumber\\
&&-[\mu_\eta^2+\beta_1^{}\phi_1^\dagger\phi_1^{}+\beta_2^{}\phi_2^\dagger\phi_2^{}+\beta_3^{}\chi^\dagger_{}\chi
\nonumber\\
&&+\beta_4^{}\textrm{Tr}(\Delta^\dagger_{}\Delta)]\eta^\dagger_{}\eta-\beta_5^{}(\eta^\dagger_{}\eta)^2_{}\,,
\end{eqnarray}
\end{subequations}
with the covariant derivative,
\begin{eqnarray}
D_\mu^{}\eta&=&\left(\partial_\mu^{}-ig\frac{\tau_a^{}}{2}W^a_\mu+i\frac{1}{2}g'B^{}_\mu\right)\eta\,.
\end{eqnarray}
The masses of the Higgs doublets $\eta$ should be
\begin{eqnarray}
M_\eta^2&\simeq&\mu_\eta^2+\beta_3^{}\langle\chi\rangle^2_{}\gg \langle\phi\rangle^2_{}\,.
\end{eqnarray}

\subsection{Type-III Dirac seesaw}

In the type-III Dirac seesaw, we have the fermion triplets with or without hypercharge, i.e.
\begin{subequations}
\begin{eqnarray}
\psi_L^{}(\!\!\begin{array}{c}1,3,0)(-1\end{array}\!\!)&=&\left[\begin{array}{cc}\frac{1}{\sqrt{2}}\psi_{L}^{0}
&\psi_{2L}^{+}\\
[2mm]
\psi_{1L}^{-}&-\frac{1}{\sqrt{2}}\psi_{L}^{0}\end{array}\right]\!,\nonumber\\
\psi'^{}_L(\!\!\begin{array}{c}1,3,0)(+2\end{array}\!\!)&=&\left[\begin{array}{cc}\frac{1}{\sqrt{2}}\psi'^{0}_{L}
&\psi'^{-}_{2L}\\
[2mm]
\psi'^{+}_{1L}&-\frac{1}{\sqrt{2}}\psi'^{0}_{L}\end{array}\right]\!;\\
[2mm]
\xi_L^{}(\!\!\begin{array}{c}1,3,+1)(-1\end{array}\!\!)&=&\left[\begin{array}{cc}\frac{1}{\sqrt{2}}\xi_{L}^{+}
&\xi_{L}^{++}\\
[2mm]
\xi_{L}^{0}&-\frac{1}{\sqrt{2}}\xi_{L}^{+}\end{array}\right]\!,\nonumber\\
\xi'^{}_L(\!\!\begin{array}{c}1,3,-1)(+2\end{array}\!\!)&=&\left[\begin{array}{cc}\frac{1}{\sqrt{2}}\xi'^{-}_{L}
&\xi'^{0}_{L}\\
[2mm]
\xi'^{--}_{L}&-\frac{1}{\sqrt{2}}\xi'^{-}_{L}\end{array}\right]\!,.
\end{eqnarray}
\end{subequations}
The kinetic and Yukawa terms are
\begin{subequations}
\begin{eqnarray}
\mathcal{L}&\supset& i\textrm{Tr}(\bar{\psi}_L^{}\gamma^\mu_{}D_\mu^{}\psi_L^{})+i\textrm{Tr}(\bar{\psi}'^{}_L\gamma^\mu_{}D_\mu^{}\psi'^{}_L)
\nonumber\\
&&-y_{\psi}^{}\bar{l}_L^{c}i\tau_2^{}\psi^{}_L\tilde{\phi}_1^{}-y_{\psi'}^{}\textrm{Tr}(\bar{\psi'}^c_Li\tau_2^{}\Sigma i\tau_2^{})\nu_R^c
\nonumber\\
&&-f_{\psi}^{}\chi\textrm{Tr}(\bar{\psi'}_L^{}i\tau_2^{}\psi^c_L i\tau_2^{})+\textrm{H.c.}\,;\\
[2mm]
\mathcal{L}&\supset& i\textrm{Tr}(\bar{\xi}_L^{}\gamma^\mu_{}D_\mu^{}\xi_L^{})+i\textrm{Tr}(\bar{\xi}'^{}_L\gamma^\mu_{}D_\mu^{}\xi'^{}_L)
\nonumber\\
&&-y_{\xi}^{}\bar{l}_L^{c}i\tau_2^{}\xi^{}_L\phi_2^{}-y_{\xi'}^{}\textrm{Tr}(\bar{\xi'}^c_L i\tau_2^{}\Delta i\tau_2^{})\nu_R^c
\nonumber\\
&&-f_{\xi}^{}\chi\textrm{Tr}(\bar{\xi'}_L^{}i\tau_2^{}\xi^c_L i\tau_2^{})+\textrm{H.c.}\,,
\end{eqnarray}
\end{subequations}
where the covariant derivatives are given by
\begin{subequations}
\begin{eqnarray}
D_\mu^{}\psi_L^{}&=&\partial_\mu^{}\psi_L^{} -ig\left[\frac{\tau_a^{}}{2}W^a_\mu,\psi_L^{}\right]\,,\nonumber\\
D_\mu^{}\psi'^{}_L&=&\partial_\mu^{}\psi'^{}_L-ig\left[\frac{\tau_a^{}}{2}W^a_\mu,\psi'^{}_L\right]\,;\\
[2mm]
D_\mu^{}\xi_L^{}&=&\partial_\mu^{}\xi_L^{} -ig\left[\frac{\tau_a^{}}{2}W^a_\mu,\xi_L^{}\right]-ig'B_\mu^{}\xi_L^{}\,,\nonumber\\
D_\mu^{}\xi'^{}_L&=&\partial_\mu^{}\xi'^{}_L-ig\left[\frac{\tau_a^{}}{2}W^a_\mu,\xi'^{}_L\right]+ig'B_\mu^{}\xi'^{}_L\,.~~~~
\end{eqnarray}
\end{subequations}

After the PQ symmetry breaking, the fermion triplets can obtain the gauge-invariant masses, i.e.
\begin{subequations}
\begin{eqnarray}
\mathcal{L}&\supset&-M_{\psi}^{}\textrm{Tr}(\bar{\psi'}_L^{}i\tau_2^{}\psi^c_L i\tau_2^{})+\textrm{H.c.}
~~\textrm{with}~~M_\psi^{}=f_\psi^{}\langle\chi\rangle\,;\nonumber\\
&&\\
\mathcal{L}&\supset&-M_{\xi}^{}\textrm{Tr}(\bar{\xi'}_L^{}i\tau_2^{}\xi^c_L i\tau_2^{})+\textrm{H.c.}
~~\textrm{with}~~M_\xi^{}=f_\xi^{}\langle\chi\rangle\,.\nonumber\\
&&
\end{eqnarray}
\end{subequations}
Without loss of generality and for convenience, we can choose a basis where the masses of the fermion triplets are real and diagonal, i.e.
\begin{subequations}
\begin{eqnarray}
M_\psi^{}&=&\textrm{diag}\{M_{\psi_1}^{},M_{\psi_2}^{},...\}\,;\\
[2mm]
M_\xi^{}&=&\textrm{diag}\{M_{\xi_1}^{},M_{\xi_2}^{},...\}\,.
\end{eqnarray}
\end{subequations}
In this basis, we can define the vector-like fermions as below,
\begin{subequations}
\begin{eqnarray}
\psi^{}_a&=&\psi^{c}_{La}+\psi'^{}_{La}\,,\\
[2mm]
\xi^{}_a&=&\xi^{c}_{La}+\xi'^{}_{La}\,.
\end{eqnarray}
\end{subequations}

\section{Neutrino masses}

In this section we will demonstrate the neutrino mass generation in the type-I, II and III Dirac seesaw models. Specifically, we will show the Dirac neutrino masses can be highly suppressed by the ratio of the constrained VEVs of the Higgs triplets over the heavy masses of the fermion singlets, the Higgs doublets or the fermion triplets.

\subsection{Neutrino masses from the type-I Dirac seesaw}

\begin{figure}
\vspace{6.7cm} \epsfig{file=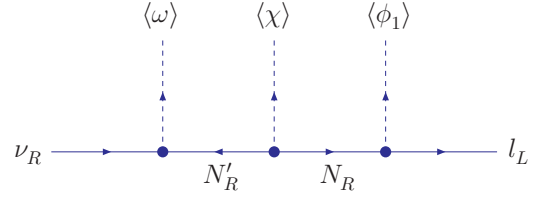, bbllx=3.5cm, bblly=6.0cm,
bburx=13.5cm, bbury=16cm, width=7cm, height=7cm, angle=0,
clip=0} \vspace{-10.75cm} \caption{\label{numass1} The Dirac neutrino masses suppressed by the heavy fermion singlets $N=N'^c_R+N_R^{}$. Here the scalar singlet $\omega$ has a VEV smaller than the VEVs of the Higgs triplets.}
\end{figure}

In the type-I Dirac seesaw model, the scalar singlet $\omega$ has a quartic coupling with two Higgs doublets and one Higgs triplet. See the $\kappa_{\omega\Sigma}^{}$-term and $\kappa_{\omega\Delta}^{}$-term in Eq. (\ref{lag1}). Accordingly, this scalar singlet can acquire an induced VEV,
\begin{subequations}
\begin{eqnarray}
\langle\omega\rangle&=&-\frac{\kappa_{\omega\Sigma}^{}\langle\sigma^0_{}\rangle\langle\phi_1^{0}\rangle\langle\phi_2^{0}\rangle}{\sqrt{2}M_\omega^2}
=-\frac{\kappa_{\omega\Sigma}^{}\langle\sigma^0_{}\rangle\langle\phi_{}^{0}\rangle^2_{}\sin2\beta}{2\sqrt{2}M_\omega^2}\nonumber\\
&=&0.031\,\textrm{GeV}\times \left(\frac{\langle\sigma^0_{}\rangle}{2.9\,\textrm{GeV}}\right)
\left(\frac{1\,\textrm{TeV}}{M_\omega^{}}\right)^2_{}\nonumber\\
&&\times\left(\frac{\kappa_{\omega\Sigma}^{}}{1}\right)\left(\frac{\sin2\beta}{1}\right)\,;\\
[2mm]
\langle\omega\rangle&=&-\frac{\kappa_{\omega\Delta}^{}\langle\delta^0_{}\rangle\langle\phi_2^{0}\rangle^2_{}}{M_\omega^2}
=-\frac{\kappa_{\omega\Delta}^{}\langle\delta^0_{}\rangle\langle\phi^0_{}\rangle^2_{}\cos^2_{}\beta}{M_\omega^2}\nonumber\\
&=&0.033\,\textrm{GeV}\times \left(\frac{\langle\delta^0_{}\rangle}{2.2\,\textrm{GeV}}\right)
\left(\frac{1\,\textrm{TeV}}{M_\omega^{}}\right)^2_{}\nonumber\\
&&\times \left(\frac{\kappa_{\omega\Delta}^{}}{1}\right)\left(\frac{\cos\beta}{1/\sqrt{2}}\right)^2_{}\,.
\end{eqnarray}
\end{subequations}
Here we have input $\langle\phi^0_{}\rangle\simeq 174\,\textrm{GeV}$.

As shown in Fig. \ref{numass1}, the type-I Dirac seesaw model then can give a tiny mass term between the left-handed neutrinos $\nu_L^{}$ and the right-handed neutrinos $\nu_R^{}$ by integrating out the heavy vector-like fermion singlets $N=N'^{c}_R+N_R^{}$,
\begin{subequations}
\begin{eqnarray}
\mathcal{L}\!&\supset&\!-m_\nu^{}\bar{\nu}_L^{}\nu_R^{}+\textrm{H.c.}~~\textrm{with}\nonumber\\
&&m_\nu^{}=-y_{N}^{}\frac{\langle\omega\rangle\langle\phi_1^{0}\rangle}{M_N^{}}y_{N'}^{}\nonumber\\
&&\quad~\,
=y_{N}^{}\frac{\kappa_{\omega\Sigma}^{}\langle\sigma^0_{}\rangle\langle\phi_{}^{0}\rangle^3_{}\sin^2_{}\beta\cos\beta}{\sqrt{2}M_\omega^2M_N^{}}y_{N'}^{}\nonumber\\
&&\quad~\,=-y_\nu^{}\langle\phi_{}^{0}\rangle =0.17\,\textrm{eV}\times \left(\frac{y_\nu^{}}{10^{-12}_{}}\right)\,;\\
[2mm]
\mathcal{L}\!&\supset&\!-m_\nu^{}\bar{\nu}_L^{}\nu_R^{}+\textrm{H.c.}~~\textrm{with}\nonumber\\
&&m_\nu^{}=-y_{N}^{}\frac{\langle\omega\rangle\langle\phi_1^{0}\rangle}{M_N^{}}y_{N'}^{}\nonumber\\
&&\quad~\,
=y_{N}^{}\frac{\kappa_{\omega\Delta}^{}\langle\delta^0_{}\rangle\langle\phi_{}^{0}\rangle^3_{}\sin\beta\cos^2_{}\beta}{M_\omega^2M_N^{}}y_{N'}^{}\nonumber\\
&&\quad~\,=-y_\nu^{}\langle\phi_{}^{0}\rangle=0.17\,\textrm{eV}\times \left(\frac{y_\nu^{}}{10^{-12}_{}}\right) \,.
\end{eqnarray}
\end{subequations}
One can easily read the effective Yukawa couplings of the right-handed neutrinos to the SM lepton and Higgs doublets,
\begin{subequations}
\begin{eqnarray}
y_\nu^{}&=&-y_{N}^{}\frac{\kappa_{\omega\Sigma}^{}\langle\sigma^0_{}\rangle\langle\phi_{}^{0}\rangle^2_{}\sin^2_{}\beta\cos\beta}
{\sqrt{2}M_\omega^2M_N^{}}y_{N'}^{}\ll1\,;\\
[2mm]
y_\nu^{}&=&-y_{N}^{}\frac{\kappa_{\omega\Delta}^{}\langle\delta^0_{}\rangle\langle\phi_{}^{0}\rangle^2_{}\sin\beta\cos^2_{}\beta}{M_\omega^2M_N^{}}y_{N'}^{}\ll1\,.
\end{eqnarray}
\end{subequations}
Since the scalar singlet is expected near the electroweak scale, the above effective Yukawa couplings can be highly suppressed by the ratio of the VEVs of the Higgs triplets over the heavy masses of the fermion singlets. For example, by inputting $M_\omega^{}=1\,\textrm{TeV}$ and $\beta=\pi/4$, we read
\begin{subequations}
\begin{eqnarray}
y_\nu^{}&=&-2.2\times 10^{-12}_{}\times
\left(\frac{y_{N}^{}}{1}\right)\left(\frac{10^{10}_{}\,\textrm{GeV}}{M_N^{}}\right)
\left(\frac{y_{N'}^{}}{1}\right)\nonumber\\
&&\times \left(\frac{\kappa_{\omega\Sigma}^{}}{1}\right)\left(\frac{\langle\sigma^0_{}\rangle}{2.9\,\textrm{GeV}}\right)\,;\\
[2mm]
y_\nu^{}&=&-2.4\times 10^{-12}_{}\times
\left(\frac{y_{N}^{}}{1}\right)\left(\frac{10^{10}_{}\,\textrm{GeV}}{M_N^{}}\right)
\left(\frac{y_{N'}^{}}{1}\right)\nonumber\\
&&\times \left(\frac{\kappa_{\omega\Delta}^{}}{1}\right)\left(\frac{\langle\delta^0_{}\rangle}{2.2\,\textrm{GeV}}\right)\,.
\end{eqnarray}
\end{subequations}

It should be noted that two or more vector-like fermion singlets are required to give two or three nonzero neutrino mass eigenvalues from the neutrino oscillation data \cite{olive2014}.

\subsection{Neutrino masses from the type-II Dirac seesaw}

In the type-II Dirac seesaw model, the Higgs doublets $\eta$ can acquire the induced VEVs after the PQ and electroweak symmetries are both broken, i.e.
\begin{subequations}
\begin{eqnarray}
\langle\eta\rangle&=&\left[\begin{array}{c}\langle\eta_{}^{0}\rangle\\
[2mm]
0\end{array}\right]~~\textrm{with}~~\langle\eta_{}^{0}\rangle\simeq -\frac{\kappa_{\eta\Sigma}^{}\langle\chi\rangle\langle\phi_{}^{0}\rangle\langle\sigma^0_{}\rangle\sin\beta}{\sqrt{2}M_\eta^2}\,;\nonumber\\
&&\\
[2mm]
\langle\eta\rangle&=&\left[\begin{array}{c}\langle\eta_{}^{0}\rangle\\
[2mm]
0\end{array}\right]~~\textrm{with}~~\langle\eta_{}^{0}\rangle\simeq -\frac{\kappa_{\eta\Delta}^{}\langle\chi\rangle\langle\phi^{0}_{}\rangle\langle\delta^0_{}\rangle\cos\beta}{M_\eta^2}\,.\nonumber\\
&&
\end{eqnarray}
\end{subequations}
Obviously, the VEVs $\langle\eta\rangle$ should be highly suppressed because of the heavy masses $M_\eta^{}$, i.e.
\begin{subequations}
\begin{eqnarray}
\langle\eta_{}^{0}\rangle\ll\langle\sigma^0_{}\rangle<\langle\phi^0_{}\rangle ~~\textrm{for}~~M_\eta^{}\gtrsim \kappa_{\eta\Sigma}^{}\langle\chi\rangle\gg\langle\phi^0_{}\rangle,\langle\sigma^0_{}\rangle\,;\nonumber\\
&&\\
[2mm]
\langle\eta_{}^{0}\rangle\ll\langle\delta^0_{}\rangle<\langle\phi^0_{}\rangle~~\textrm{for}~~M_\eta^{}\gtrsim \kappa_{\eta\Delta}^{}\langle\chi\rangle\gg\langle\phi^0_{}\rangle,\langle\delta^0_{}\rangle\,.\nonumber\\
&&
\end{eqnarray}
\end{subequations}

\begin{figure}
\vspace{8.0cm} \epsfig{file=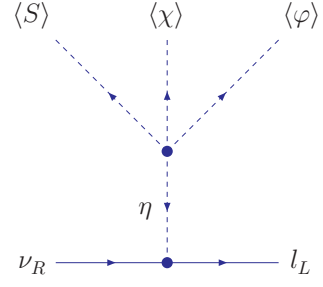, bbllx=3.5cm, bblly=6.0cm,
bburx=13.5cm, bbury=16cm, width=7cm, height=7cm, angle=0,
clip=0} \vspace{-10.75cm} \caption{\label{numass2} The Dirac neutrino masses suppressed by the heavy Higgs doublets $\eta$. Here $(\varphi,S)$ stands for $(\tilde{\phi}_1^{},\Sigma)$ or $(\phi_2^{},\Delta)$.}
\end{figure}

Through their Yukawa interactions with the heavy Higgs doublets $\eta$, the left-handed neutrinos $\nu_L^{}$ and the right-handed neutrinos $\nu_R^{}$ then can obtain a tiny Dirac mass term,
\begin{subequations}
\label{numasss}
\begin{eqnarray}
\mathcal{L}&\supset&-m_\nu^{}\bar{\nu}_L^{}\nu_R^{}+\textrm{H.c.}~~\textrm{with}\nonumber\\
&&m_\nu^{}=y_\eta^{}\langle\eta\rangle=-y_\eta^{}
\frac{\kappa_{\eta\Sigma}^{}\langle\chi\rangle\langle\phi_{}^{0}\rangle\langle\sigma^0_{}\rangle\sin\beta}{\sqrt{2}M_\eta^2}\nonumber\\
&&\quad~\, =y_\nu^{}\langle\phi_{}^{0}\rangle=0.17\,\textrm{eV}\times \left(\frac{y_\nu^{}}{10^{-12}_{}}\right)\,;\\
[2mm]
\mathcal{L}&\supset&-m_\nu^{}\bar{\nu}_L^{}\nu_R^{}+\textrm{H.c.}~~\textrm{with}\nonumber\\
&&m_\nu^{}=y_\eta^{}\langle\eta\rangle=-y_\eta^{}
\frac{\kappa_{\eta\Delta}^{}\langle\chi\rangle\langle\phi_{}^{0}\rangle\langle\delta^0_{}\rangle\cos\beta}{M_\eta^2}\nonumber\\
&&\quad~\, =y_\nu^{}\langle\phi_{}^{0}\rangle=0.17\,\textrm{eV}\times \left(\frac{y_\nu^{}}{10^{-12}_{}}\right)\,.
\end{eqnarray}
\end{subequations}
Here we have introduced the effective Yukawa couplings of the right-handed neutrinos to the SM lepton and Higgs doublets as
\begin{subequations}
\begin{eqnarray}
y_\nu^{}&=&-y_\eta^{}
\frac{\kappa_{\eta\Sigma}^{}\langle\chi\rangle\langle\sigma^0_{}\rangle\sin\beta}{\sqrt{2}M_\eta^2}\nonumber\\
&=&-1.5\times 10^{-12}_{}\times \left(\frac{\langle\sigma^0_{}\rangle}{2.9\,\textrm{GeV}}\right)\left(\frac{10^{12}_{}\,\textrm{GeV}}{M_\eta^{}}\right)^2_{}\nonumber\\
&&\times \left(\frac{\langle\chi\rangle}{10^{12}_{}\,\textrm{GeV}}\right)\left(\frac{y_{\eta}^{}}{1}\right)\left(\frac{\kappa_{\eta\Sigma}^{}}{1}\right)
\left(\frac{\sin\beta}{1/\sqrt{2}}\right)\,;~~~\\
[2mm]
y_\nu^{}&=&-y_\eta^{}
\frac{\kappa_{\eta\Delta}^{}\langle\chi\rangle\langle\delta^0_{}\rangle\cos\beta}{M_\eta^2}\nonumber\\
&=&-1.6\times 10^{-12}_{}\times \left(\frac{\langle\delta^0_{}\rangle}{2.2\,\textrm{GeV}}\right)\left(\frac{10^{12}_{}\,\textrm{GeV}}{M_\eta^{}}\right)^2_{}\nonumber\\
&&\times \left(\frac{\langle\chi\rangle}{10^{12}_{}\,\textrm{GeV}}\right)\left(\frac{y_{\eta}^{}}{1}\right)\left(\frac{\kappa_{\eta\Delta}^{}}{1}\right)
\left(\frac{\cos\beta}{1/\sqrt{2}}\right)\,,~~~
\end{eqnarray}
\end{subequations}
which can be highly suppressed by the ratio of the VEVs of the Higgs triplets over the masses of the heavy Higgs doublets. This scheme of the Dirac neutrino mass generation can also be understood by Fig. \ref{numass2}, where we denoted $(\tilde{\phi}_1^{},\Sigma)$ and $(\phi_2^{},\Delta)$ by $(\varphi,S)$.

\subsection{Neutrino masses from the type-III Dirac seesaw}

\begin{figure}
\vspace{6.5cm} \epsfig{file=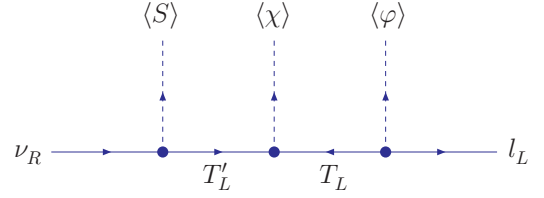, bbllx=3.5cm, bblly=6.0cm,
bburx=13.5cm, bbury=16cm, width=7cm, height=7cm, angle=0,
clip=0} \vspace{-10.75cm} \caption{\label{numass3} The Dirac neutrino masses suppressed by the heavy fermion triplets $T=T_L^{c}+T'^{}_L$. Here $(T^{}_{L},T'^{}_{L},\varphi,S)$ stands for $(\psi^{}_{L},\psi'^{}_{L},\tilde{\phi}_1^{},\Sigma)$ or $(\xi^{}_{L},\xi'^{}_{L},\phi_2^{},\Delta)$.}
\end{figure}

In the type-III Dirac seesaw model, we can integrate out the heavy vector-like fermion triplets to induce a mass term between the left-handed neutrinos and the right-handed neutrinos, i.e.
\begin{subequations}
\label{numassf}
\begin{eqnarray}
\mathcal{L}&\supset&-m_\nu^{}\bar{\nu}_L^{}\nu_R^{}+\textrm{H.c.}~~\textrm{with}\nonumber\\
&&m_\nu^{}=-y_{\psi}^{\ast}\frac{\langle\sigma^0_{}\rangle\langle\phi_1^{0}\rangle}{\sqrt{2}M_\psi^{}}y_{\psi'}^{\ast}
=-y_{\psi}^{\ast}\frac{\langle\sigma^0_{}\rangle\langle\phi_{}^{0}\rangle\sin\beta}{\sqrt{2}M_\psi^{}}y_{\psi'}^{\ast}\nonumber\\
&&\quad~\,=-y_\nu^{}\langle\phi_{}^{0}\rangle=-0.17\,\textrm{eV}\times \left(\frac{y_\nu^{}}{10^{-12}_{}}\right) \,;\\
[2mm]
\mathcal{L}&\supset&-m_\nu^{}\bar{\nu}_L^{}\nu_R^{}+\textrm{H.c.}~~\textrm{with}\nonumber\\
&&m_\nu^{}=-y_{\xi}^{\ast}\frac{\langle\delta^0_{}\rangle\langle\phi_2^{0}\rangle}{M_\xi^{}}y_{\xi'}^{\ast}
=-y_{\xi}^{\ast}\frac{\langle\delta^0_{}\rangle\langle\phi_{}^{0}\rangle\cos\beta}{M_\xi^{}}y_{\xi'}^{\ast}\nonumber\\
&&\quad~\,=-y_\nu^{}\langle\phi_{}^{0}\rangle =-0.17\,\textrm{eV}\times \left(\frac{y_\nu^{}}{10^{-12}_{}}\right)\,.
\end{eqnarray}
\end{subequations}
Here we have introduced the effective Yukawa couplings of the right-handed neutrinos to the SM lepton and Higgs doublets as
\begin{subequations}
\begin{eqnarray}
y_\nu^{}&=&y_{\psi}^{\ast}\frac{\langle\sigma^0_{}\rangle\sin\beta}{\sqrt{2}M_\psi^{}}y_{\psi'}^{\ast}\nonumber\\
&=&1.3\times 10^{-12}_{}\times \left(\frac{y_{\psi}^{\ast}}{0.3}\right)\left(\frac{10^{11}_{}\,\textrm{GeV}}{M_\psi^{}}\right)
\left(\frac{y_{\psi'}^{\ast}}{0.3}\right)\nonumber\\
&&\times \left(\frac{\langle\sigma^0_{}\rangle}{2.9\,\textrm{GeV}}\right)
\left(\frac{\sin\beta}{1/\sqrt{2}}\right)\,;\\
[2mm]
y_\nu^{}&=&y_{\xi}^{\ast}\frac{\langle\delta^0_{}\rangle\cos\beta}{M_\xi^{}}y_{\xi'}^{\ast}\nonumber\\
&=&1.4\times 10^{-12}_{}\times \left(\frac{y_{\xi}^{\ast}}{0.3}\right)\left(\frac{10^{11}_{}\,\textrm{GeV}}{M_\xi^{}}\right)
\left(\frac{y_{\xi'}^{\ast}}{0.3}\right)\nonumber\\
&&\times \left(\frac{\langle\delta^0_{}\rangle}{2.2\,\textrm{GeV}}\right)
\left(\frac{\cos\beta}{1/\sqrt{2}}\right)\,.
\end{eqnarray}
\end{subequations}
The relevant diagram is shown in Fig. \ref{numass3} where $(T^{}_{L},T'^{}_{L},\varphi,S)$ stands for $(\psi^{}_{L},\psi'^{}_{L},\tilde{\phi}_1^{},\Sigma)$ or $(\xi^{}_{L},\xi'^{}_{L},\phi_2^{},\Delta)$. Clearly the effective Yukawa couplings can be highly suppressed by the ratio of the VEVs of the Higgs triplets over the masses of the heavy vector-like fermion triplets.

Note we need two or more vector-like fermion triplets to give two or three nonzero neutrino mass eigenvalues required by the neutrino oscillation data \cite{olive2014}.

\subsection{Neutrino masses from the combined Dirac seesaw models}

The neutrino masses can also be induced by the combined type-I+II, type-I+III, type-II+III or type-I+II+III Dirac seesaw models. These combined models can give two or three nonzero neutrino mass eigenvalues even if they only contain one vector-like fermion singlets and/or one vector-like fermion triplets.

\section{Baryon asymmetry}

In this section we will illustrate how to generate the cosmic baryon asymmetry in the type-I, II and III Dirac seesaw models. Specifically, a lepton asymmetry stored in the SM left-handed leptons and an opposite lepton asymmetry stored in the right-handed neutrinos can be produced in the CP-violating and out-of-equilibrium decays of the heavy fermion singlets, Higgs doublets or fermion triplets. The related masses and couplings of these heavy fields are also responsible for generating the light Dirac neutrino masses. Since (i) the right-handed neutrinos do not participate in the $SU(2)_L^{}$ sphalerons, (ii) the effective Yukawa interactions between the left- and right-handed neutrinos go into equilibrium at a very low temperature where the sphalerons have stopped working, the right-handed neutrino asymmetry will not affect the baryon asymmetry, instead, only the left-handed lepton asymmetry will be partially converted to the baryon asymmetry.

\subsection{Neutrinogenesis in the type-I Dirac seesaw}

\begin{figure*}
\vspace{5.5cm} \epsfig{file=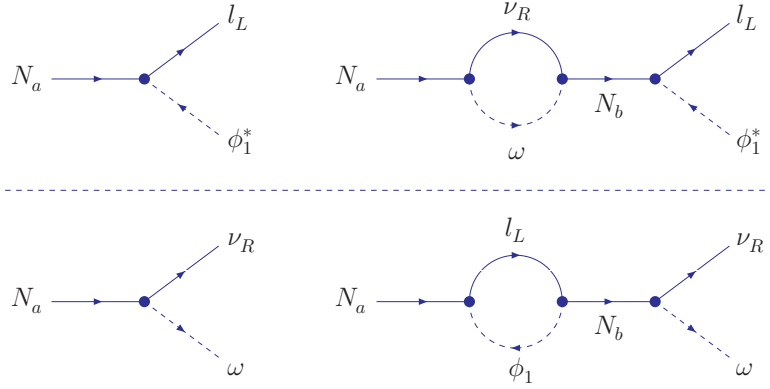, bbllx=4.5cm, bblly=6.0cm,
bburx=14.5cm, bbury=16cm, width=7cm, height=7cm, angle=0,
clip=0} \vspace{-7cm} \caption{\label{decay1} The lepton-number-conserving decays of the heavy fermion singlets $N=N'^c_R+N_R^{}$.}
\end{figure*}

In the type-I Dirac seesaw model, the vector-like fermion singlets $N=N_R^{}+N'^{c}_R$ can have the two-body decays as shown in Fig. \ref{decay1}. We calculate the decay widths at tree level,
\begin{eqnarray}
\Gamma_{N_a}^{}&=&\Gamma(N_a^{}\rightarrow l_L^{}+\phi_1^\ast)+\Gamma(N_a^{}\rightarrow \nu_R^{}+\omega)\nonumber\\
&=&\Gamma(N_a^{c}\rightarrow l_L^{c}+\phi_{1}^{})+\Gamma(N_a^{c}\rightarrow \nu_R^c+\omega^\ast_{})\nonumber\\
&=&\frac{1}{16\pi}\left[(y_{N}^\dagger y_{N}^{})_{aa}^{}+\frac{1}{2}(y_{N'}^{} y_{N'}^{\dagger})_{aa}^{}\right]M_{N_a}^{}\,,
\end{eqnarray}
and the CP asymmetries at one-loop level,
\begin{eqnarray}
\!\!\!\!\!\!\!\!\varepsilon_{N_a}^{}&=&\frac{\Gamma(N_a^{}\rightarrow l_L^{}+\phi_1^\ast)-\Gamma(N_a^{c}\rightarrow l_L^{c}+\phi^{}_{1})}{\Gamma_{N_a}^{}}\nonumber\\
\!\!\!\!\!\!\!\!&=&\frac{\Gamma(N_a^{c}\rightarrow \nu_R^c+\omega^\ast_{})-\Gamma(N_a^{}\rightarrow \nu_R^{}+\omega)}{\Gamma_{N_a}^{}}\nonumber\\
\!\!\!\!\!\!\!\!&=&\frac{1}{8\pi}\frac{\textrm{Im}\left[(y_{N}^\dagger y_{N}^{})_{ab}^{}(y_{N'}^{} y_{N'}^{\dagger})_{ba}\right]}
{(y_{N}^\dagger y_{N}^{})_{aa}^{}+\frac{1}{2}(y_{N'}^{} y_{N'}^{\dagger})_{aa}^{}}\frac{M_{N_a}^{}M_{N_b}^{}}{M_{N_b}^{2}-M_{N_a}^{2}}\,.\nonumber\\
\!\!\!\!\!\!\!\!&&
\end{eqnarray}
The final baryon asymmetry then can be given by \cite{kt1990}
\begin{eqnarray}
\eta_B^{}&=&\frac{n_B^{}}{n_\gamma^{}}=7.04\times \frac{n_B^{}}{s}=7.04\times \left(-\frac{28}{79}\right) \times \frac{n_L^{}}{s}\nonumber\\
&=& 7.04\times \left(-\frac{28}{79}\right)\times\frac{\sum_{a}^{}\varepsilon_{N_a}^{}r_{N_a}^{}}{g_\ast^{}}\,.
\end{eqnarray}
Here and thereafter $n_B^{}$, $n_L^{}$, $n_\gamma^{}$ and $s$ are the baryon number density, the lepton number density, the photon number density and the entropy density, respectively. The factor $-\frac{28}{79}$ is the sphaleron lepton-to-baryon coefficient. The washout coefficient $r_{N_a}^{}\leq 1$ can be determined by the related Boltzmann equations. The relativistic degrees of freedom during the leptogenesis epoch, $g_{\ast}^{}$, can be given by $g_\ast^{}=106.75+4+2+3=115.75$ (The SM fields plus one Higgs doublet, one complex Higgs singlet and one real Higgs triplet) or $g_\ast^{}=106.75+4+2+6=118.75$ (The SM fields plus one Higgs doublet, one complex Higgs singlet and one complex Higgs triplet).

Note that in the above calculations we have assumed the initial sate $N_a^{}$ is much heavier than the $\phi'$ fraction of the final state $\phi_{1}^{}$. If the $\phi'$ fraction is heavier than the decaying fields, only the $\phi$ fraction of the final states $\phi_{1}^{}$ will contribute to the corresponding decay widths and CP asymmetries. This case will not be studied in details here and thereafter. The definition of the $\phi$ and $\phi'$ scalars can be found in Eq. (\ref{phi}).

Instead of deriving and then numerically solving the Boltzmann equations, we adopt an analytical approximation \cite{kt1990} to give the final baryon asymmetry. For this purpose, we assume a hierarchical spectrum of the fermion singlets $N_{1,2,...}^{}$. Consequently, the final baryon asymmetry should come from the decays of the lightest fermion singlet denoted by $N_1^{}$. For demonstration, we define
\begin{eqnarray}
K_{N_1^{}}^{}&=&\frac{\Gamma_{N_1^{}}^{}}{2H(T)}\left|_{T=M_{N_1^{}}^{}}^{}\right.\,,
\end{eqnarray}
where $H(T)$ is the Hubble constant,
\begin{eqnarray}
H=\left(\frac{8\pi^{3}_{}g_{\ast}^{}}{90}\right)^{\frac{1}{2}}_{}
\frac{T^{2}_{}}{M_{\textrm{Pl}}^{}}\,,
\end{eqnarray}
with $M_{\textrm{Pl}}^{}=1.22\times 10^{19}_{}\,\textrm{GeV}$ being the Planck mass. For $1\ll K_{N_1^{}}^{} \lesssim 10^6_{}$, the final baryon asymmetry can well approximate to \cite{kt1990}
\begin{eqnarray}
\label{ba1}
\eta_B^{}&=&7.04\times \left(-\frac{28}{79}\right)\times \frac{\varepsilon_{N_1^{}}^{}}{g_\ast^{}K_{N_1^{}}^{} z_{N_1^{}}^{}} \nonumber\\
&&\textrm{with}~~z_{N_1^{}}^{}=\frac{M_{N_1^{}}^{}}{T_{N_1^{}}^{}}\simeq 4.2(\ln K_{N_1^{}}^{})^{0.6}_{}\,.
\end{eqnarray}
We then simply take,
\begin{eqnarray}
y_{N'}^{}=y_N^T\,,
\end{eqnarray}
so that we can parametrize,
\begin{eqnarray}
y_N^{}&=&i\frac{U \sqrt{\hat{m}_\nu^{}}O\sqrt{M_N^{}}}{\sqrt{\langle\omega\rangle\langle\phi^0_{}\rangle\sin\beta}} ~~\textrm{with}\nonumber\\
[2mm]
&& m_\nu^{}=U \hat{m}_\nu^{} U^T_{}=U\textrm{diag}\{m_1^{}\,,~m_2^{}\,,~m_3^{}\}U^T_{}\,, \nonumber\\
[2mm]
&&OO^T_{}=O^T_{}O=1\,,
\end{eqnarray}
and then derive
\begin{eqnarray}
\label{pa1}
\varepsilon_{N_1}^{}&<&\varepsilon_{N_1}^{\textrm{max}}=\frac{1}{12\,\pi}\frac{M_{N_1}^{}m_{\textrm{max}}}{\langle\omega\rangle\langle\phi^0_{}\rangle\sin\beta}\,,\nonumber\\
[2mm]
K_{N_1}^{}&=&\frac{3}{32\,\pi}\left(\frac{90}{8\pi^{3}_{}g_{\ast}^{}}\right)^{\frac{1}{2}}_{}
\frac{M_{\textrm{Pl}}^{}\tilde{m}_{1}^{}}{\langle\omega\rangle\langle\phi^0_{}\rangle\sin\beta}~~\textrm{with}\nonumber\\
[2mm]
&&m_{\textrm{max}}^{}=\max \{m_1^{}\,,~m_2^{}\,,~m_3^{}\}\,,\nonumber\\
[2mm]
&&m_{\textrm{min}}^{}=\min \{m_1^{}\,,~m_2^{}\,,~m_3^{}\}\,,\nonumber\\
[2mm]
&&\tilde{m}_i^{}\equiv (O^\dagger_{}\hat{m}_\nu^{}O)_{ii}^{}\in (m_{\textrm{min}}^{}\,,~m_{\textrm{max}}^{})\,.
\end{eqnarray}
By inputting
\begin{eqnarray}
&&M_{N_1^{}}^{}=10^{9}_{}\,\textrm{GeV}\,,~~\langle\omega\rangle=0.03\,\textrm{GeV}\,,~~\sin\beta=\frac{1}{\sqrt{2}}\,,\nonumber\\
&&m_{\textrm{max}}^{}=0.05\,\textrm{eV}\,,~~\tilde{m}_{1}^{}=10^{-4}_{}\,\textrm{eV}\,,
\end{eqnarray}
in Eqs. (\ref{ba1}) and (\ref{pa1}), we read
\begin{eqnarray}
\varepsilon_{N_1^{}}^{\textrm{max}}&=&3.6\times 10^{-4}_{}\,,\nonumber\\
K_{N_1^{}}^{}&=&575\times \left(\frac{106.75}{g_\ast^{}}\right)^{\frac{1}{2}}_{}\,,\nonumber\\
z_{N_1^{}}^{}&=&12.7\times \left(1+\frac{\ln\sqrt{106.75/g_\ast^{}}}{\ln575}\right)^{0.6}_{}\,.
\end{eqnarray}
The final baryon asymmetry then can be consistent to the observation \cite{olive2014},
\begin{eqnarray}
\eta_B^{}=6.3\times 10^{-10}_{}\times \left(\frac{\varepsilon_{N_1^{}}^{}}{-0.55\,\varepsilon_{N_1^{}}^{\textrm{max}}}\right) \left(\frac{106.75}{g_\ast^{}}\right)^{\frac{1}{2}}_{}\,.
\end{eqnarray}

\subsection{Neutrinogenesis in the type-II Dirac seesaw}

\begin{figure*}
\vspace{5.5cm} \epsfig{file=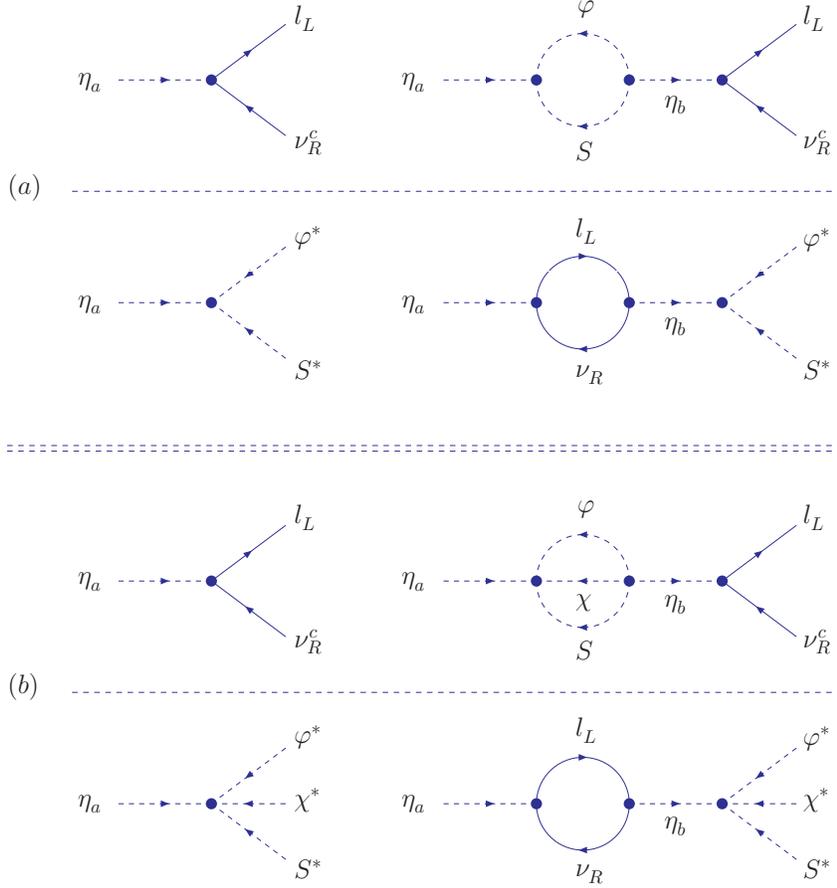, bbllx=4.5cm, bblly=6.0cm,
bburx=14.5cm, bbury=16cm, width=7cm, height=7cm, angle=0,
clip=0} \vspace{-0.5cm} \caption{\label{decay2} The lepton-number-conserving decays of the heavy Higgs doublets $\eta$. Here $(\varphi,S)$ stands for $(\tilde{\phi}_1^{},\Sigma)$ or $(\phi_2^{},\Delta)$.}
\end{figure*}

In the type-II Dirac seesaw model, the Higgs doublets $\eta$ can be lighter than the PQ symmetry breaking scale $\langle\chi\rangle$. In this case, the two-body decays of the Higgs doublets $\eta$, as shown in Fig. \ref{decay2}a, can generate a lepton asymmetry stored in the left-handed leptons $l_L^{}$ and an opposite lepton asymmetry stored in the right-handed neutrinos $\nu_R^{}$ as long as the CP is not conserved. The decay widths at tree level are
\begin{eqnarray}
\Gamma_{\eta_a}^{}&=&\Gamma(\eta_a^{}\rightarrow l_L^{}+\nu_R^c)+\Gamma(\eta_a^{}\rightarrow S^\ast_{}+\varphi^\ast_{})\nonumber\\
&=&\Gamma(\eta_a^{\ast}\rightarrow l_L^{c}+\nu_R^{})+\Gamma(\eta_a^{\ast}\rightarrow S+\varphi)\nonumber\\
&=&\frac{1}{16\pi}\left[\textrm{Tr}(y_{\eta_a}^\dagger y_{\eta_a}^{})+\frac{3 \kappa^2_{\eta_a S}\langle\chi\rangle^2_{}}{2 M_{\eta_a}^{2}}\right]M_{\eta_a}^{}\,,
\end{eqnarray}
while the CP asymmetries at one-loop order are
\begin{eqnarray}
\label{cpa2s}
\varepsilon_{\eta_a}^{S}&=&\frac{\Gamma(\eta_a^{}\rightarrow l_L^{}+\nu_R^c)-\Gamma(\eta_a^{\ast}\rightarrow l_L^{c}+\nu_R^{})}{\Gamma_{\eta_a}^{}}\nonumber\\
&=&\frac{\Gamma(\eta_a^{\ast}\rightarrow S+\varphi)-\Gamma(\eta_a^{}\rightarrow S^\ast_{}+\varphi^\ast_{})}{\Gamma_{\eta_a}^{}}\nonumber\\
&=&\frac{3}{8\pi}\frac{\textrm{Im}[\textrm{Tr}(y_{\eta_a}^\dagger y_{\eta_b}^{})]}
{\textrm{Tr}(y_{\eta_a}^\dagger y_{\eta_a}^{})+\frac{3 \kappa^2_{\eta_a S}\langle\chi\rangle^2_{}}{2 M_{\eta_a}^{2}}}\frac{\kappa_{\eta_a S}^{}\kappa_{\eta_b S}^{}\langle\chi\rangle^2_{}}{M_{\eta_b}^{2}-M_{\eta_a}^{2}}\,.
\end{eqnarray}
Alternatively, the Higgs doublets $\eta$ can have their heavy masses before the PQ symmetry breaking. We then should consider the two-body and three-body decays as shown in Fig. \ref{decay2}b. The decay widths should be
\begin{eqnarray}
\Gamma_{\eta_a}^{}&=&\Gamma(\eta_a^{}\rightarrow l_L^{}+\nu_R^c)+\Gamma(\eta_a^{}\rightarrow S^\ast_{}+\varphi^\ast_{}+\chi^\ast_{})\nonumber\\
&=&\Gamma(\eta_a^{\ast}\rightarrow l_L^{c}+\nu_R^{})+\Gamma(\eta_a^{\ast}\rightarrow S+\varphi+\chi)\nonumber\\
&=&\frac{1}{16\pi}\left[\textrm{Tr}(y_{\eta_a}^\dagger y_{\eta_a}^{})+\frac{3\kappa^2_{\eta_a S}}{64\pi^2_{}}\right]M_{\eta_a}^{}\,.
\end{eqnarray}
As for the CP asymmetries, they should be
\begin{eqnarray}
\label{cpa3s}
\varepsilon_{\eta_a}^{S}&=&\frac{\Gamma(\eta_a^{}\rightarrow l_L^{}+\nu_R^c)-\Gamma(\eta_a^{\ast}\rightarrow l_L^{c}+\nu_R^{})}{\Gamma_{\eta_a}^{}}\nonumber\\
&=&\frac{\Gamma(\eta_a^{\ast}\rightarrow S+\varphi+\chi)-\Gamma(\eta_a^{}\rightarrow S^\ast_{}+\varphi^\ast_{}+\chi^\ast_{})}{\Gamma_{\eta_a}^{}}\nonumber\\
&=&\frac{3}{256\pi^3_{}}\frac{\kappa_{\eta_a S}^{}\kappa_{\eta_b S}^{}\textrm{Im}[\textrm{Tr}(y_{\eta_a}^\dagger y_{\eta_b}^{})]}
{\textrm{Tr}(y_{\eta_a}^\dagger y_{\eta_a}^{})+\frac{3 \kappa^2_{\eta_a S}}{64\pi^2_{}}}\frac{M_{\eta_a}^{2}}{M_{\eta_b}^{2}-M_{\eta_a}^{2}}\,.
\end{eqnarray}
We emphasize at least two heavy Higgs doublets $\eta$ should be introduced to induce a nonzero CP asymmetry (\ref{cpa2s}) or (\ref{cpa3s}).

The final baryon asymmetry then can be described by \cite{kt1990}
\begin{eqnarray}
\label{bas}
\eta_B^{}&=&\frac{n_B^{}}{n_\gamma^{}}=7.04\times \frac{n_B^{}}{s} =7.04\times  \left(-\frac{28}{79}\right)\times \frac{n_L^{}}{s}\nonumber\\
&=& 7.04\times  \left(-\frac{28}{79}\right)\times \frac{\sum_{a}^{}\varepsilon_{\eta_a}^{S}r_{\eta_a}^{}}{g_\ast^{}}\times 2\,.
\end{eqnarray}
Here the factor $2$ appears because the decaying particle $\eta_{a}^{}$ is a doublet. As for the parameter $r_{\eta_a}^{}\leq 1$, it is a washout effect depending on the decay, inverse decay, scattering and annihilation involving the heavy Higgs doublets $\eta_a^{}$. Ones can derive and solve the Boltzmann equations to exactly determine the values of the coefficient $\kappa_{\eta_a}^{}$ for the given masses and couplings of the Higgs doublets $\eta_{a}^{}$. The detailed Boltzmann equations and their numerical solutions will be studied elsewhere. Instead, we adopt an analytical approximation \cite{kt1990} for demonstration. For this purpose, we assume the type-II Dirac seesaw model contains two heavy Higgs doublets $\eta_{1,2}^{}$. The final baryon asymmetry thus should be produced by the decays of the lighter Higgs doublet denoted by $\eta_1^{}$. As an example, we consider the two-body decays. In this case, the relativistic degrees of freedom $g_{\ast}^{}$ can be given by $g_\ast^{}=106.75+4+3=113.75$ (The SM fields plus one Higgs doublet and one real Higgs triplet) or $g_\ast^{}=106.75+4+6=116.75$ (The SM fields plus one Higgs doublet and one complex Higgs triplet). By setting
\begin{subequations}
\label{pa2}
\begin{eqnarray}
&&M_{\eta_1^{}}^{}=0.3\,M_{\eta_2^{}}^{}=10^{12}_{}\,\textrm{GeV}\,,~~\langle\chi\rangle=10^{12}_{}\,\textrm{GeV}\,,\nonumber\\
&&\kappa_{\eta_{1}^{} \Sigma}^{}=0.3\,\kappa_{\eta_{2}^{} \Sigma}^{}=0.6\,,~~y_{\eta_{1}^{} }^{}=y_{\eta_{2}^{}}^{}e^{i\alpha}_{}\,,\nonumber\\
&&\langle\sigma^0_{}\rangle=2.9\,\textrm{GeV}\,,~~\sin\beta=\frac{1}{\sqrt{2}}\,;\\
[2mm]
&&M_{\eta_1^{}}^{}=0.3\,M_{\eta_2^{}}^{}=10^{12}_{}\,\textrm{GeV}\,,~~\langle\chi\rangle=10^{12}_{}\,\textrm{GeV}\,,\nonumber\\
&&\kappa_{\eta_{1}^{} \Delta}^{}=0.3\,\kappa_{\eta_{2}^{} \Delta}^{}=0.6\,,~~y_{\eta_{1}^{} }^{}=y_{\eta_{2}^{}}^{}e^{i\alpha}_{}\,,\nonumber\\
&&\langle\delta^0_{}\rangle=2.2\,\textrm{GeV}\,,~~\cos\beta=\frac{1}{\sqrt{2}}\,,
\end{eqnarray}
\end{subequations}
into Eq. (\ref{numasss}), we determine
\begin{eqnarray}
&&y_{\eta_1^{}}^{}=0.7\times \left(\frac{m_\nu^{}}{0.1\,\textrm{eV}}\right)\,,\nonumber\\
&&\textrm{Tr}(y_{\eta_1^{}}^{\dagger}y_{\eta_1^{}}^{})=0.49\times \left[\frac{\textrm{Tr}(m^\dagger_\nu m_\nu^{})}{0.01\,\textrm{eV}^2}\right]\,.
\end{eqnarray}
For the parameter choice (\ref{pa2}) and the input $\textrm{Tr}(y_{\eta_1^{}}^{\dagger}y_{\eta_1^{}}^{})=0.49$, we obtain
\begin{eqnarray}
K_{\eta_1^{}}^{}&=&\frac{\Gamma_{\eta_1^{}}^{}}{2H(T)}\left|_{T=M_{\eta_1^{}}^{}}^{}\right.=7244\times \left(\frac{106.75}{g_\ast^{}}\right)^{\frac{1}{2}}_{}\,,\nonumber\\
[2mm]
z_{\eta_1^{}}^{}&=&\frac{M_{\eta_1^{}}^{}}{T_{\eta_1^{}}^{}}\simeq 4.2(\ln K_{\eta_1^{}}^{})^{0.6}_{}\nonumber\\
&=&15.6\times \left(1+\frac{\ln\sqrt{106.75/g_\ast^{}}}{\ln7244}\right)^{0.6}_{}\,,\nonumber\\
[2mm]
\varepsilon_{\eta_1^{}}^{}&=&-0.0015\times \left(\frac{\sin\alpha}{0.22}\right)\,.
\end{eqnarray}
The final baryon asymmetry thus can arrive at a desired value \cite{olive2014},
\begin{eqnarray}
\label{ba2}
\eta_B^{}&=&7.04\times \left(-\frac{28}{79}\right)\times \frac{\varepsilon_{\eta_1^{}}^{}}{g_\ast^{}K_{\eta_1^{}}^{} z_{\eta_1^{}}^{}} \times 2\nonumber\\
&=&6.2\times 10^{-10}_{}\times \left(\frac{\sin\alpha}{0.22}\right)\left(\frac{106.75}{g_\ast^{}}\right)^{\frac{1}{2}}_{}\,.
\end{eqnarray}

\subsection{Neutrinogenesis in the type-III Dirac seesaw}

\begin{figure*}
\vspace{5.5cm} \epsfig{file=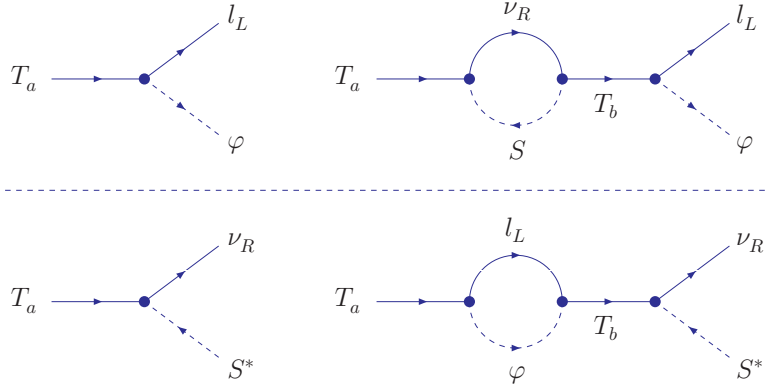, bbllx=4.5cm, bblly=6.0cm,
bburx=14.5cm, bbury=16cm, width=7cm, height=7cm, angle=0,
clip=0} \vspace{-7cm} \caption{\label{decay3} The lepton-number-conserving decays of the heavy fermion triplets $T=T_L^{c}+T'^{}_L$. Here $(T,\varphi,S)$ stands for $(\psi,\tilde{\phi}_1^{},\Sigma)$ or $(\xi,\phi_2^{},\Delta)$.}
\end{figure*}

In the type-III Dirac seesaw model, the vector-like fermion triplets $\psi=\psi^{c}_L+\psi'^{}_L$ or $\xi=\xi^{c}_L+\xi'^{}_L$ can have the two-body decays as shown in Fig. \ref{decay3}. The decay widths are given by
\begin{eqnarray}
\Gamma_{T_a}^{}&=&\Gamma(T_a^{}\rightarrow l_L^{}+\varphi)+\Gamma(T_a^{}\rightarrow \nu_R^{}+S^\ast)\nonumber\\
&=&\Gamma(T_a^{c}\rightarrow l_L^{c}+\varphi^\ast_{})+\Gamma(T_a^{c}\rightarrow \nu_R^c+S)\nonumber\\
&=&\frac{1}{32\pi}\left[(y_{T}^T y_{T}^{\ast})_{aa}^{}+(y_{T'}^\ast y_{T'}^{T})_{aa}^{}\right]M_{T_a}^{}\,,
\end{eqnarray}
while the CP asymmetries are
\begin{eqnarray}
\label{cpa2f}
\!\!\!\!\!\!\!\!\varepsilon_{T_a}^{S}&=&\frac{\Gamma(T_a^{}\rightarrow l_L^{}+\varphi)-\Gamma(T_a^{c}\rightarrow l_L^{c}+\varphi^\ast_{})}{\Gamma_{T_a}^{}}\nonumber\\
\!\!\!\!\!\!\!\!&=&\frac{\Gamma(T_a^{c}\rightarrow \nu_R^c+S)-\Gamma(T_a^{}\rightarrow \nu_R^{}+S^\ast)}{\Gamma_{T_a}^{}}\nonumber\\
\!\!\!\!\!\!\!\!&=&\frac{1}{8\pi}\frac{\textrm{Im}\left[(y_{T}^T y_{T}^{\ast})_{ab}^{}(y_{T'}^\ast y_{T'}^{T})_{ba}^{}\right]}
{(y_{T}^T y_{T}^{\ast})_{aa}^{}+(y_{T'}^\ast y_{T'}^{T})_{aa}^{}}\frac{M_{T_a}^{}M_{T_b}^{}}{M_{T_b}^{2}-M_{T_a}^{2}}\,.\nonumber\\
\!\!\!\!\!\!\!\!&&
\end{eqnarray}
Here and thereafter we denote $\psi$ and $\xi$ by $T$ in the formula. The final baryon asymmetry then can be given by \cite{kt1990}
\begin{eqnarray}
\label{baf}
\eta_B^{}&=&\frac{n_B^{}}{n_\gamma^{}}=7.04\times \frac{n_B^{}}{s} \nonumber\\
&=& 7.04\times \left(-\frac{28}{79}\right)\times\frac{\sum_{a}^{}\varepsilon_{T_a}^{S}r_{T_a}^{}}{g_\ast^{}}\times  3\,.
\end{eqnarray}
Here the factor $3$ appears because the decaying particle $T_{a}^{}$ is a triplet. The washout coefficient $r_{T_a}^{}\leq 1$ can be determined by the Boltzmann equations which will be discussed elsewhere. In the following, we consider an analytical approximation \cite{kt1990} for demonstration. As for the relativistic degrees of freedom $g_{\ast}^{}$, it can be given by $g_\ast^{}=106.75+4+3=113.75$ (The SM fields plus one Higgs doublet and one real Higgs triplet) or $g_\ast^{}=106.75+4+6=116.75$ (The SM fields plus one Higgs doublet and one complex Higgs triplet).

We assume the fermion triplets $T_{1,2,...}^{}$ have a hierarchical spectrum. Therefore, the decays of the lightest fermion triplet denoted by $T_1^{}$ should dominate the final baryon asymmetry. We then define
\begin{eqnarray}
K_{T_1^{}}^{}&=&\frac{\Gamma_{T_1^{}}^{}}{2H(T)}\left|_{T=M_{T_1^{}}^{}}^{}\right.\,.
\end{eqnarray}
For $1\ll K_{T_1^{}}^{} \lesssim 10^6_{}$, the final baryon asymmetry can well approximate to \cite{kt1990}
\begin{eqnarray}
\label{ba3}
\eta_B^{}&=&7.04\times \left(-\frac{28}{79}\right)\times \frac{\varepsilon_{T_1^{}}^{}}{g_\ast^{}K_{T_1^{}}^{} z_{T_1^{}}^{}} \times 3\nonumber\\
&&\textrm{with}~~z_{T_1^{}}^{}=\frac{M_{T_1^{}}^{}}{T_{T_1^{}}^{}}\simeq 4.2(\ln K_{T_1^{}}^{})^{0.6}_{}\,.
\end{eqnarray}
We further simply take,
\begin{eqnarray}
y_{T'}^{}=y_T^T\,,
\end{eqnarray}
so that we can parametrize,
\begin{subequations}
\begin{eqnarray}
y_\psi^{T}&=&i\frac{U \sqrt{\hat{m}_\nu^{}}O\sqrt{M_\psi^{}}}{\sqrt{\langle \sigma^0_{} \rangle\langle\phi^0_{}\rangle\sin\beta/\sqrt{2}}}\,;\\
[2mm]
y_\xi^{T}&=&i\frac{U \sqrt{\hat{m}_\nu^{}}O\sqrt{M_\xi^{}}}{\sqrt{\langle \delta^0_{} \rangle\langle\phi^0_{}\rangle\cos\beta}}\,,
\end{eqnarray}
\end{subequations}
and then derive
\begin{subequations}
\label{pa3}
\begin{eqnarray}
\varepsilon_{\psi_1}^{}&<&\varepsilon_{\psi_1}^{\textrm{max}}=\frac{1}{16\,\pi}\frac{M_{\psi_1}^{}m_{\textrm{max}}}{\langle \sigma^0_{}\rangle\langle\phi^0_{}\rangle\sin\beta/\sqrt{2}}\,,\nonumber\\
[2mm]
K_{\psi_1}^{}&=&\frac{1}{16\,\pi}\left(\frac{90}{8\pi^{3}_{}g_{\ast}^{}}\right)^{\frac{1}{2}}_{}
\frac{M_{\textrm{Pl}}^{}\tilde{m}_{1}^{}}{\langle \sigma^0_{}\rangle\langle\phi^0_{}\rangle\sin\beta/\sqrt{2}}\,;\\
[2mm]
\varepsilon_{\xi_1}^{}&<&\varepsilon_{\xi_1}^{\textrm{max}}=\frac{1}{16\,\pi}\frac{M_{\xi_1}^{}m_{\textrm{max}}}{\langle \delta^0_{}\rangle\langle\phi^0_{}\rangle\cos\beta}\,,\nonumber\\
[2mm]
K_{\xi_1}^{}&=&\frac{1}{16\,\pi}\left(\frac{90}{8\pi^{3}_{}g_{\ast}^{}}\right)^{\frac{1}{2}}_{}
\frac{M_{\textrm{Pl}}^{}\tilde{m}_{1}^{}}{\langle \delta^0_{}\rangle\langle\phi^0_{}\rangle\cos\beta}\,.
\end{eqnarray}
\end{subequations}
By inputting
\begin{subequations}
\begin{eqnarray}
&&M_{\psi_1^{}}^{}=10^{11}_{}\,\textrm{GeV}\,,~~\langle \sigma^0_{}\rangle=2.9\,\textrm{GeV}\,,~~\sin\beta=\frac{1}{\sqrt{2}}\,,\nonumber\\
&&m_{\textrm{max}}^{}=0.07\,\textrm{eV}\,,~~\tilde{m}_{1}^{}=0.001\,\textrm{eV}\,;\\
[2mm]
&&M_{\xi_1^{}}^{}=10^{11}_{}\,\textrm{GeV}\,,~~\langle \delta^0_{}\rangle=2.2\,\textrm{GeV}\,,~~\cos\beta=\frac{1}{\sqrt{2}}\,,\nonumber\\
&&m_{\textrm{max}}^{}=0.07\,\textrm{eV}\,,~~\tilde{m}_{1}^{}=0.001\,\textrm{eV}\,
\end{eqnarray}
\end{subequations}
in Eqs. (\ref{ba3}) and (\ref{pa3}), we have
\begin{eqnarray}
\varepsilon_{T_1^{}}^{\textrm{max}}&=&5.5\times 10^{-4}_{}\,,\nonumber\\
[2mm]
K_{T_1^{}}^{}&=&56\times \left(\frac{106.75}{g_\ast^{}}\right)^{\frac{1}{2}}_{}\,,\nonumber\\
[2mm]
z_{T_1^{}}^{}&=&9.7\times \left(1+\frac{\ln\sqrt{106.75/g_\ast^{}}}{\ln28}\right)^{0.6}_{}\,,
\end{eqnarray}
and then obtain an expected final baryon asymmetry \cite{olive2014},
\begin{eqnarray}
\eta_B^{}=6.4\times 10^{-10}_{}\times \left(\frac{\varepsilon_{T_1^{}}^{}}{-0.009\,\varepsilon_{T_1^{}}^{\textrm{max}}}\right)\left(\frac{106.75}{g_\ast^{}}\right)^{\frac{1}{2}}_{}\,.
\end{eqnarray}

\subsection{Neutrinogenesis in the type-II+III Dirac seesaw}

\begin{figure*}
\vspace{5.5cm} \epsfig{file=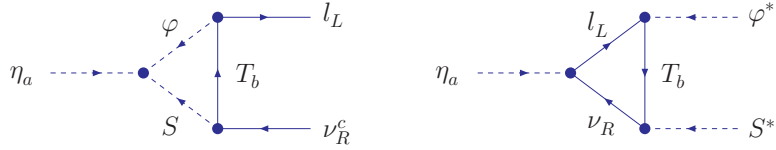, bbllx=4.5cm, bblly=6.0cm,
bburx=14.5cm, bbury=16cm, width=7cm, height=7cm, angle=0,
clip=0} \vspace{-10cm} \caption{\label{decay4} The vertex corrections in the two-body decays of the heavy Higgs doublets $\eta$. Here $(T,\varphi,S)$ stands for $(\psi,\tilde{\phi}_1^{},\Sigma)$ or $(\xi,\phi_2^{},\Delta)$.}
\end{figure*}

In the combined model of the type-II+III Dirac seesaw, we have not only the heavy Higgs doublets but also the heavy fermion triplets. The CP asymmetries in the decays of the heavy Higgs doublets then should contain the vertex corrections as below,
\begin{eqnarray}
\label{cpavs}
\varepsilon_{\eta_a}^{V}&=&\frac{\Gamma(\eta_a^{}\rightarrow l_L^{}+\nu_R^c)-\Gamma(\eta_a^{\ast}\rightarrow l_L^{c}+\nu_R^{})}{\Gamma_{\eta_a}^{}}\nonumber\\
&=&\frac{\Gamma(\eta_a^{\ast}\rightarrow S+\varphi)-\Gamma(\eta_a^{}\rightarrow S^\ast_{}+\varphi^\ast_{})}{\Gamma_{\eta_a}^{}}\nonumber\\
&=&\frac{3}{8\pi}\frac{\textrm{Im}[(y_{T'}^\ast y_{\eta_a}^\dagger y_{T}^\ast)_{bb}^{}]}{\textrm{Tr}(y_{\eta_a}^\dagger y_{\eta_a}^{})+\frac{3\kappa_{\eta_a S}^2\langle\chi\rangle^2_{}}{2M_{\eta_a}^2}}\frac{\kappa_{\eta_a S}^{}\langle\chi\rangle M_{T_b}^{}}{M_{\eta_a}^2}\nonumber\\
&&\times\ln\left(1+\frac{M_{\eta_a}^2}{M_{T_b}^{2}}\right)\,.
\end{eqnarray}
The relevant diagrams are shown in Fig. \ref{decay4}. Note such vertex corrections can appear till the fermion triplets obtain their heavy masses after the PQ symmetry breaking. We can conveniently treat the total CP asymmetries in the decays of the heavy Higgs doublets by
\begin{eqnarray}
\label{cpa2vs}
\varepsilon_{\eta_a}^{}=\varepsilon_{\eta_a}^{S}+\varepsilon_{\eta_a}^{V}\,,
\end{eqnarray}
where the self-energy correction $\varepsilon_{\eta_a}^{S}$ is given by Eq. (\ref{cpa2s}). The final baryon asymmetry (\ref{bas}) from the decays of the heavy Higgs doublets then should be replaced by
\begin{eqnarray}
\eta_B^{}&=& 7.04\times  \left(-\frac{28}{79}\right)\times \frac{\sum_{a}^{}\varepsilon_{\eta_a}^{}r_{\eta_a}^{}}{g_\ast^{}}\times 2\,.
\end{eqnarray}
If only one heavy Higgs doublet is introduced to the models, we will have no self-energy corrections (\ref{cpa2s}) to the CP asymmetries (\ref{cpa2vs}). In this case, the vertex corrections (\ref{cpavs}) should be the unique source to the CP asymmetries (\ref{cpa2vs}). Otherwise, both the vertex corrections (\ref{cpavs}) and the self-energy corrections (\ref{cpa2s}) should be taken into account.

\begin{figure*}
\vspace{5.5cm} \epsfig{file=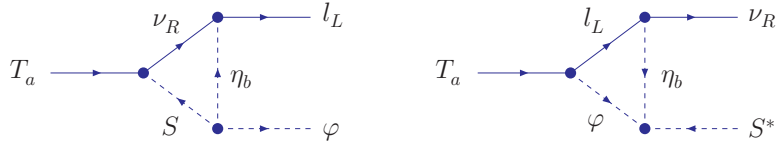, bbllx=4.5cm, bblly=6.0cm,
bburx=14.5cm, bbury=16cm, width=7cm, height=7cm, angle=0,
clip=0} \vspace{-10cm} \caption{\label{decay5} The vertex corrections in the two-body decays of the heavy fermion triplets $T=T_L^{c}+T'^{}_L$. Here $(T,\varphi,S)$ stands for $(\psi,\tilde{\phi}_1^{},\Sigma)$ or $(\xi,\phi_2^{},\Delta)$.}
\end{figure*}

Similarly, from Fig. \ref{decay5}, we can compute the vertex corrections to the CP asymmetries in the decays of the heavy fermion triplets,
\begin{eqnarray}
\label{cpavf}
\varepsilon_{T_a}^{V}&=&\frac{\Gamma(T_a^{}\rightarrow l_L^{}+\varphi)-\Gamma(T_a^{c}\rightarrow l_L^{c}+\varphi^\ast_{})}{\Gamma_{T_a}^{}}\nonumber\\
&=&\frac{\Gamma(T_a^{c}\rightarrow \nu_R^c+S)-\Gamma(T_a^{}\rightarrow \nu_R^{}+S^\ast)}{\Gamma_{T_a}^{}}\nonumber\\
&=&\frac{1}{4\pi}\frac{\textrm{Im}\left[(y_{T}^T y_{\eta_b}^{}y_{T'}^T )_{aa}^{}\right]}
{(y_{T}^T y_{T}^{\ast})_{aa}^{}+(y_{T'}^\ast y_{T'}^{T})_{aa}^{}}\frac{\kappa_{\eta_b S}^{}\langle\chi\rangle}{M_{T_a}^{}}\nonumber\\
&&\times \left[1-\frac{M_{\eta_b}^{2}}{M_{T_a}^2}\ln\left(1+\frac{M_{T_a}^2}{M_{\eta_b}^{2}}\right)\right]\,.
\end{eqnarray}
The total CP asymmetries in the decays of the heavy fermion triplets can be conveniently given by
\begin{eqnarray}
\label{cpa2vf}
\varepsilon_{T_a}^{}=\varepsilon_{T_a}^{S}+\varepsilon_{T_a}^{V}\,.
\end{eqnarray}
Accordingly, the final baryon asymmetry (\ref{baf}) from the decays of the heavy fermion triplets should be replaced by \cite{kt1990}
\begin{eqnarray}
\eta_B^{}&=& 7.04\times  \left(-\frac{28}{79}\right)\times \frac{\sum_{a}^{}\varepsilon_{T_a}^{}r_{T_a}^{}}{g_\ast^{}}\times 3\,.
\end{eqnarray}
In the case with two or more heavy fermion triplets and at least one heavy Higgs doublet, the CP asymmetries (\ref{cpa2vf}) should contain not only the vertex corrections (\ref{cpavf}) but also the self-energy corrections (\ref{cpa2f}). If the model contains only one heavy fermion triplet, the vertex corrections (\ref{cpavf}) rather than the self-energy corrections (\ref{cpa2f}) should be the source of the CP asymmetries (\ref{cpa2vf}).

\section{Summary}

In this paper we have proposed a class of Dirac seesaw-leptogenesis models to simultaneously explain the small Dirac neutrino masses and the cosmic matter-antimatter asymmetry. Our models contain three right-handed neutrinos, an iso-triplet Higgs scalar, as well as some heavy gauge-singlet fermions, iso-doublet Higgs scalars and/or iso-triplet fermions, besides the fields in the DSFZ invisible axion model. The iso-triplet fields can carry a zero or nonzero hypercharge. After the PQ and electroweak symmetry breaking, the effective Yukawa couplings of the Dirac neutrinos to the SM Higgs scalar can be highly suppressed by the ratio of the VEVs of the iso-triplet Higgs scalars over the masses of the gauge-singlet fermions, the iso-doublet Higgs scalars or the iso-triplet fermions. The PQ symmetry breaking scale can be constrained by the invisible axion while the VEVs of the iso-triplet Higgs scalars can be constrained by the $\rho$ parameter. Through the decays of the heavy gauge-singlet fermions, iso-doublet Higgs scalars or iso-triplet fermions, we can obtain a lepton asymmetry in the left-handed leptons and an opposite lepton asymmetry in the right-handed neutrinos. Since the right-handed neutrinos do not participate in the sphaleron processes, the left-handed lepton asymmetry can be partially converted to a baryon asymmetry. The axion can serve as a dark matter particle as it is in the DFSZ model.

\textbf{Acknowledgement}: I would like to thank Hong-Jian He for helpful discussions. This work was supported by the Recruitment Program for Young Professionals under Grant No. 15Z127060004, the Shanghai Jiao Tong University under Grant No. WF220407201 and the Shanghai Laboratory for Particle Physics and Cosmology under Grant No. 11DZ2260700.


\begin{thebibliography}{99}



\bibitem{olive2014}
K.A. Olive {\it et al.}, (Particle Data Group Collaboration), Chin. Phys. C \textbf{38}, 090001 (2014).



\bibitem{minkowski1977}
P. Minkowski, Phys. Lett. B \textbf{67}, 421 (1977); T. Yanagida, in
{\it Proceedings of the Workshop on Unified Theory and the Baryon
Number of the Universe}, edited by O. Sawada and A. Sugamoto (KEK,
Tsukuba, 1979), p. 95; M. Gell-Mann, P. Ramond, and R. Slansky, in
{\it Supergravity}, edited by F. van Nieuwenhuizen and D. Freedman
(North Holland, Amsterdam, 1979), p. 315; S.L. Glashow, in {\it
Quarks and Leptons}, edited by M. L\'{e}vy {\it et al.} (Plenum, New
York, 1980), p. 707; R.N. Mohapatra and G. Senjanovi\'{c}, Phys.
Rev. Lett. \textbf{44}, 912 (1980).


\bibitem{mw1980}
M. Magg and C. Wetterich, Phys. Lett. B \textbf{94}, 61 (1980); J.
Schechter and J.W.F. Valle, Phys. Rev. D \textbf{22}, 2227 (1980);
T.P. Cheng and L.F. Li, Phys. Rev. D \textbf{22}, 2860 (1980); G.
Lazarides, Q. Shafi, and C. Wetterich, Nucl. Phys. B \textbf{181},
287 (1981); R.N. Mohapatra and G. Senjanovi\'{c}, Phys. Rev. D
\textbf{23}, 165 (1981).


\bibitem{flhj1989}
R. Foot, H. Lew, X.G. He, and G.C. Joshi, Z. Phys. C \textbf{44},
441 (1989).


\bibitem{ma1998}
E. Ma, Phys. Rev. Lett. \textbf{81}, 1171 (1998).


\bibitem{barr2003}
S.M. Barr, Phys. Rev. Lett. \textbf{92}, 101601 (2004).




\bibitem{krs1985}
V.A. Kuzmin, V.A. Rubakov, and M.E. Shaposhnikov, Phys. Lett. B
\textbf{155}, 36 (1985).

\bibitem{fy1986}
M. Fukugita and T. Yanagida, Phys. Lett. B \textbf{174}, 45 (1986).



\bibitem{lpy1986}
P. Langacker, R.D. Peccei, and T. Yanagida, Mod. Phys. Lett. A
\textbf{1}, 541 (1986); M.A. Luty, Phys. Rev. D \textbf{45}, 455
(1992); R.N. Mohapatra and X. Zhang, Phys. Rev. D \textbf{46}, 5331 (1992).




\bibitem{fps1995}
M. Flanz, E.A. Paschos, and U. Sarkar, Phys. Lett. B \textbf{345},
248 (1995); M. Flanz, E.A. Paschos, U. Sarkar, and J. Weiss, Phys.
Lett. B \textbf{389}, 693 (1996); L. Covi, E. Roulet, and F.
Vissani, Phys. Lett. B \textbf{384}, 169 (1996); A. Pilaftsis, Phys.
Rev. D \textbf{56}, 5431 (1997).


\bibitem{ms1998}
E. Ma and U. Sarkar, Phys. Rev. Lett. \textbf{80}, 5716 (1998).


\bibitem{bcst1999}
R. Barbieri, P. Creminelli, A. Strumia, and N. Tetradis, Nucl. Phys. B \textbf{575}, 61 (2000).




\bibitem{hambye2001}
T. Hambye, Nucl. Phys. B \textbf{633}, 171 (2002).


\bibitem{di2002}
S. Davidson and A. Ibarra, Phys. Lett. B \textbf{535}, 25 (2002); W.
Buchm\"{u}ller, P. Di Bari, and M. Pl\"{u}macher, Nucl. Phys. B
\textbf{665}, 445 (2003).

\bibitem{gnrrs2003}
G.F. Giudice, A. Notari, M. Raidal, A. Riotto, and A. Strumia, Nucl. Phys. B \textbf{685}, 89 (2004).


\bibitem{hs2004}
T. Hambye and G. Senjanovi\'{c}, Phys. Lett. B \textbf{582}, 73
(2004); S. Antusch and S.F. King, Phys. Lett. B \textbf{597}, 199
(2004); P. Gu and X.J. Bi, Phys. Rev. D \textbf{70}, 063511 (2004).


\bibitem{bbp2005}
W. Buchmuller, P. Di Bari, and M. Plumacher, Annals Phys. \textbf{315}, 305 (2005).



\bibitem{ma2006}
E. Ma, Phys. Rev. D \textbf{73}, 077301 (2006);
E. Ma, Annales Fond. Broglie \textbf{31}, 285 (2006).



\bibitem{dnn2008}
S. Davidson, E. Nardi, and Y. Nir, Phys. Rept. \textbf{466}, 105
(2008).


\bibitem{dhh2014}
F.F. Deppisch, J. Harz, and M. Hirsch, Phys. Rev. Lett. \textbf{112}, 221601 (2014).

\bibitem{ksy2015}
A. Kusenko, K. Schmitz, T.T. Yanagida, Phys. Rev. Lett. \textbf{115}, 011302 (2015).

\bibitem{fmmn2015}
S. Blanchet, P.S. Bhupal Dev, and R.N. Mohapatra, Phys. Rev. D \textbf{82}, 115025 (2010)
P.S. Bhupal Dev, P. Millington, A. Pilaftsis, and D. Teresi, Nucl. Phys. B \textbf{886}, 569 (2014);
C.S. Fong, D. Meloni, A. Meroni, and E. Nardi, JHEP \textbf{1501}, 111 (2015);
P.S. Bhupal Dev, P. Millington, A. Pilaftsis, and D. Teresi, Nucl. Phys. B \textbf{891}, 128 (2015);
P. Di Bari, L. Marzola, M. Re Fiorentin, Nucl. Phys. B \textbf{893}, 122 (2015);
J.D. Clarke, R. Foot, R.R. Volkas, Phys. Rev. D \textbf{91}, 073009 (2015);
J. Zhang, Phys. Rev. D \textbf{91}, 073012 (2015);
S. Lavignac and B. Schmauch, JHEP \textbf{1505}, 124 (2015);
M. Dhuria, C. Hati, R. Rangarajan, and U. Sarkar, Phys. Rev. D \textbf{92}, 031701 (2015);
M. Ibe and K. Kaneta, Phys. Rev. D \textbf{92}, 035019 (2015);
P.S. Bhupal Dev, P. Millington, A. Pilaftsis, and D. Teresi, Nucl. Phys. B \textbf{897}, 749 (2015);
L. Pearce, L. Yang, A. Kusenko, M. Peloso, Phys. Rev. D \textbf{92}, 023509 (2015);
S. Kashiwase, H. Okada, Y. Orikasa, and T. Toma, arXiv:1505.04665 [hep-ph];
F. Bj\"{o}rkeroth, F.J. de Anda, I. de Medeiros Varzielas, and S.F. King, JHEP \textbf{1510}, 104 (2015);
J.D. Clarke, R. Foot, R.R. Volka, Phys. Rev. D \textbf{92}, 033006 (2015);
M. Aoki, N. Haba, and R. Takahashi, PTEP \textbf{2015}, 113B03 (2015);
A. Pilaftsis, D. Teresi, Phys. Rev. D \textbf{92}, 085016 (2015);
A. Abada, G. Arcadi, V. Domcke, and M. Lucente, JCAP \textbf{1511}, 041 (2015);
P. Di Bari and S.F. King, JCAP \textbf{1510} 008 (2015);
P. Hern\'{o}ndez, M. Kekic, J. L\'{o}pez-Pav\'{o}n, J. Racker, and N. Rius, JHEP \textbf{1510}, 067 (2015);
J.I. McDonald and G.M. Shore, Phys. Lett. B \textbf{751}, 469 (2015);
R. Kalita and D. Borah, Phys. Rev. D \textbf{92}, 055012 (2015);
T. Ishihara, N. Maekawa, M. Takegawa, and M. Yamanaka, JHEP \textbf{1602}, 108 (2016);
J. Gehrlein, S.. Petcov, M. Spinrath, and X. Zhang, Nucl. Phys. B \textbf{899}, 617 (2015);
B. Karmakar and A. Sil, Phys. Rev. D \textbf{93}, 013006 (2016);
A. Addazi, M. Bianchi, and G. Ricciardi, JHEP \textbf{1602}, 035 (2016);
K.J. Bae, H. Baer, H. Serce, and Y.F. Zhang, JCAP \textbf{1601}, 012 (2016);
J.M. Cline, A. Diaz-Furlong, and J. Ren, Phys. Rev. D \textbf{93}, 036009 (2016);
E.T. Franco, Phys. Rev. D \textbf{92}, 113010 (2015);
C. Hati and U. Sarkar, arXiv:1511.02874 [hep-ph].




\bibitem{rw1983}
M. Roncadelli and D. Wyler, Phys. Lett. B \textbf{133}, 325 (1983);
P. Roy and O. Shanker, Phys. Rev. Lett. \textbf{52}, 713 (1984).




\bibitem{dlrw1999}
K. Dick, M. Lindner, M. Ratz, and D. Wright, Phys. Rev. Lett.
\textbf{84}, 4039 (2000).

\bibitem{mp2002}
H. Murayama and A. Pierce, Phys. Rev. Lett. \textbf{89}, 271601
(2002).

\bibitem{gh2006}
P.H. Gu and H.J. He, JCAP \textbf{0612}, 010 (2006).



\bibitem{tt2006}
B. Thomas and M. Toharia, Phys. Rev. D \textbf{73}, 063512 (2006);
S. Abel and V. Page, JHEP \textbf{0605}, 024 (2006);
P.H. Gu, H.J. He, and U. Sarkar, JCAP \textbf{0711}, 016 (2007);
E.J. Chun and P. Roy, JHEP \textbf{0806}, 089 (2008);
P.H. Gu, H.J. He, and U. Sarkar, Phys. Lett. B \textbf{659}, 634 (2008);
P.H. Gu and U. Sarkar, Phys. Rev. D \textbf{77}, 105031 (2008);
A. Bechinger and G. Seidl, Phys. Rev. D \textbf{81}, 065015 (2010);
H. Davoudiasl and I. Lewis, Phys. Rev. D \textbf{86}, 015024 (2012);
K.Y. Choi, E.J. Chun, and C.S. Shin, Phys. Lett. B \textbf{723}, 90 (2013);
P.H. Gu, arXiv:1410.5753 [hep-ph];
P.H. Gu, JCAP \textbf{1412}, 046 (2014);
P.H. Gu and X.G. He, arXiv:1511.03835 [hep-ph].





\bibitem{dsz2007}
F. del Aguila, J. Syska, M. Zralek, Phys. Rev. D \textbf{76}, 013007 (2007);
C. Luhn and M. Thormeier, Phys. Rev. D \textbf{77}, 056002 (2008);
Y. Farzan and E. Ma, Phys. Rev. D \textbf{86}, 033007 (2012);
X.w. Liu and S. Zhou, Int. J. Mod. Phys. A \textbf{28}, 1350040 (2013);
M. Drewes, Int. J. Mod. Phys. E \textbf{22}, 1330019 (2013);
G. Abbas, S. Gupta, G. Rajasekaran, and R. Srivastava, Phys. Rev. D \textbf{91}, 111301 (2015);
H. Okada, arXiv:1404.0280 [hep-ph];
A. de Gouv\^{e}a, D. Hern\'{a}ndez, JHEP \textbf{1510}, 046 (2015);
A. Esmaili and A. Yu. Smirnov, Phys. Rev. D \textbf{92}, 093012 (2015).




\bibitem{gu2012}
P.H. Gu, Nucl. Phys. B \textbf{872}, 38 (2013).

\bibitem{kim1979}
J.E. Kim, Phys. Rev. Lett. \textbf{43}, 103 (1979); M.A. Shifman,
A.I. Vainshtein, and V.I. Zakharov, Nucl. Phys. B \textbf{166}, 493
(1980).



\bibitem{dfs1981}
M. Dine, W. Fischler, and M. Srednicki, Phys. Lett. B \textbf{104},
199 (1981); A.R. Zhitnitsky, Sov. J. Nucl. Phys. \textbf{31}, 260
(1980).



\bibitem{pq1977}
R.D. Peccei and H.R. Quinn, Phys. Rev. Lett. \textbf{38}, 1440
(1977); Phys. Rev. D \textbf{16}, (1977).

\bibitem{weinberg1978}
S. Weinberg, Phys. Rev. Lett. \textbf{40}, 223 (1978).

\bibitem{wilczek1978}
F. Wilczek, Phys. Rev. Lett. \textbf{40}, 279 (1978).



\bibitem{shin1987}
M. Shin, Phys. Rev. Lett. \textbf{59}, 2515 (1987); X.G. He and R.R.
Volkas, Phys. Lett. B \textbf{208}, 261 (1988); C.Q. Geng and J.N.
Ng, Phys. Rev. D \textbf{39}, 1925 (1989); C.Q. Geng and J.N. Ng,
Mod. Phys. Lett. A \textbf{4}, 581 (1989); C.Q. Geng and J.N. Ng,
Phys. Rev. D \textbf{39}, 1449 (1989); Z.G. Berezhiani and M.Yu.
Khlopov, Z. Phys. C \textbf{49}, 73 (1991); D.A. Demir, E. Ma, and
U. Sarkar, J. Phys. G \textbf{26}, L117 (2000); E. Ma, Phys. Lett. B
\textbf{514}, 330 (2001); E. Ma, J. Phys. G \textbf{29}, 313 (2003);
E. Ma, Mod. Phys. Lett. A \textbf{22}, 2721 (2007);
P.H. Gu and M. Lindner, Phys. Lett. B \textbf{698}, 40 (2011);
P.H. Gu and M. Lindner, Phys. Lett. B \textbf{697}, 229 (2011);
C.S Chen and L.H. Tsai, Phys. Rev. D \textbf{88}, 055015 (2013).





\bibitem{kt1990}
E.W. Kolb and M.S. Turner, \textit{The Early Universe},
Addison-Wesley, 1990.


\end{thebibliography}
\end{document}